\documentclass[10pt,twocolumn,twoside]{IEEEtran}
\IEEEoverridecommandlockouts

\def\BibTeX{{\rm B\kern-.05em{\sc i\kern-.025em b}\kern-.08em
    T\kern-.1667em\lower.7ex\hbox{E}\kern-.125emX}}
\usepackage{easyReview}
\usepackage{textcomp}
\usepackage{xcolor}
\usepackage{tikz}
\usepackage{cite,nicefrac}
\usepackage[ruled]{algorithm2e}
\SetKwFor{For}{for }{do}{{end for}}
\usepackage{graphicx}
\usepackage{amsmath,amssymb,amsfonts}
\usepackage{bm}
\usepackage[version=4]{mhchem}
\usepackage{siunitx}
\usepackage{longtable,tabularx}
\usepackage{setspace}
\usepackage{subcaption}
\usepackage{array}
\usepackage{pdfpages}
\usepackage{lipsum}
\usepackage{wrapfig}
\usepackage{hyperref}
\usepackage{stmaryrd}
\usepackage{pifont}
\usepackage[stable]{footmisc}
\usepackage{stackengine} 
\usepackage{authblk}
\usepackage{mathtools}
\usepackage[utf8]{inputenc}
\usepackage{soul}
\usepackage{url}
\usepackage{mathtools}
\usepackage{amssymb}
\usepackage{amsmath,amsfonts,bbold}
\usepackage{textcomp}
\usepackage{stfloats}
\usepackage{verbatim}
\usepackage{bbm}
 \usepackage{mdframed}
\usepackage{amsthm}
\usepackage{multirow}
\usepackage{setspace}

\newtheorem{lemma}{Lemma}

\newtheorem{theorem}{Theorem}
\newtheorem{Def}{Definition}

\newtheorem{assumption}{Assumption}

\def\OmegaVec{{\bm\varOmega}_\text{vec}}
\def\OmegaC{{\bm\varOmega}_\text{vec}^\complement}

\newcommand{\mathbm}[1]{\bm{\mathcal{#1}}}
\newcommand{\barbm}[1]{\bar{\bm{#1}}}
\newcommand{\tildebm}[1]{\tilde{\bm{#1}}}
\newcommand{\hattext}[1]{\widehat{\text{#1}}}

\DeclareMathOperator*{\minimize}{\textrm{minimize}}

\DeclareMathOperator*{\argmin}{arg\,min}

\definecolor{mygreen}{RGB}{0, 128, 0}
\definecolor{mygreen2}{RGB}{0, 200, 128}

\allowdisplaybreaks

\begin{document}

\title{Radio Map Estimation via Latent Domain Plug-and-Play Denoising 
}

\author{Le~Xu, Lei~Cheng, Junting~Chen, Wenqiang~Pu, and Xiao~Fu
\thanks{L. Xu and X. Fu are with the School of Electrical Engineering and
Computer Science, Oregon State University, OR, USA. (email: xul2@oregonstate.edu, xiao.fu@oregonstate.edu).}
\thanks{L. Cheng is with the College of Information Science and Electronic Engineering, Zhejiang University, Hangzhou, China. (email: lei\_cheng@zju.edu.cn).}
\thanks{J. Chen is with the School of Science and Engineering, and Shenzhen Future Network of Intelligence Institute (FNii-Shenzhen), The Chinese University of Hong Kong, Shenzhen, China. (email: juntingc@cuhk.edu.cn)}
\thanks{W. Pu is with the Shenzhen Research Institute of Big Data, The Chinese University of Hong Kong, Shenzhen, China. (email: wpu@sribd.cn)}

\thanks{
The work of L. Cheng was supported in part by the National Natural Science Foundation of China (NSFC) under
Grant No. 62371418.
The work of J. Chen was supported in part by 
NSFC under
Grant No. 62171398.
The work of L. Xu and X. Fu was supported in part by the National Science Foundation (NSF) under Project NSF ECCS-2024058.
The work of X. Fu was also supported in part by NSF under project NSF CCF-2210004.}
}

\maketitle

\begin{abstract}
Radio map estimation (RME), also known as \textit{spectrum cartography}, aims to reconstruct the strength of radio interference across different domains (e.g., space and frequency) from sparsely sampled measurements. 
To tackle this typical inverse problem, state-of-the-art RME methods rely on handcrafted or data-driven structural information of radio maps.
However, the former often struggles to model complex radio frequency (RF) environments and the latter requires excessive training---making it hard to quickly adapt to {\it in situ} sensing tasks.
This work presents a spatio-spectral RME approach based on {\it plug-and-play} (PnP) denoising, a technique from computational imaging.
The idea is to leverage the observation that the denoising operations of signals like natural images and radio maps are similar---despite the nontrivial differences of the signals themselves.
Hence, sophisticated denoisers designed for or learned from natural images can be directly employed to assist RME, avoiding using
radio map data for training.
Unlike conventional PnP methods that operate directly in the data domain, 
the proposed method exploits the underlying physical structure of radio maps and proposes an ADMM algorithm that denoises in a latent domain. This design significantly improves computational efficiency and enhances noise robustness.
Theoretical aspects, e.g., recoverability of the complete radio map and convergence of the ADMM algorithm
are analyzed.
Synthetic and real data experiments are conducted to demonstrate the effectiveness of our approach.
\end{abstract}

\begin{IEEEkeywords}
Radio map estimation, ADMM, plug-and-play denoising, tensor completion, recoverability analysis
\end{IEEEkeywords}

\section{Introduction}
\label{sec:introduction}
Radio map estimation (RME), also known as spectrum cartography, aims to construct a map of the received signal indicators (e.g., the strength of radio interference) across different domains (e.g., space and frequency) from sparsely acquired measurements.
RME plays
a crucial role in promoting radio frequency (RF) environmental awareness. It is also considered a key enabler for more intelligent and efficient use of spectrum resources; see \cite{zeng2024tutorial, Bi2019radiomap,romero2022radio}.

From a signal processing viewpoint, RME presents an ill-posed inverse problem.
Like other inverse problems, RME boils down to imposing proper structural constraints onto the target radio map to ensure accurate recovery.
Earlier RME methods rely on manually crafted structural constraints, e.g., smoothness, sparsity \cite{Bazerque2011Splines,kim2010cooperative,Boccolini2012Kriging} and low matrix/tensor rank \cite{Zhang2020SpectrumViaBlockTerm,sun2024integrated,Sun2022InterpolationMatrix}. 
Nonetheless, these constraints may not always hold in practice, especially under complex RF environments, e.g., when heavy shadowing exists.
Recent advances have introduced data-driven---especially deep learning based---structural constraints. Leveraging the expressive power of neural networks, these methods learn to effectively represent detailed characteristics of radio maps
\cite{Shrestha2022DeepSpectrum,Subash2024Quantized,Levie2021radiounet,teganya2021deep,roger2023deep,romero2022radio}. However, neural network-based structural representations require a large amount of (historical or simulator-generated) radio map data for training, which is often unavailable for new or quickly changing environments.

A notable development for inverse problem solving is the so-called \textit{plug-and-play} (PnP) denoising framework \cite{Kamilov2023PNP,venkatakrishnan2013plug}, which was popularized in the computational imaging community. 
The PnP method leverages sophisticated or deeply learned natural image denoisers to solve inverse problems for other types of data, e.g., ocean sound speed field (SSF) data \cite{li2024zero}, hyperspectral image (HSI) data \cite{Liu2022HSIpnp}, and magnetic resonance imaging (MRI) data \cite{wei2020tuning,liu2020rare}.
The rationale of PnP denoising is that many types of data (e.g., natural images and radio maps)---despite having different data characteristics---share similar denoising processes. This is because denoising boils down to removing erratic perturbations from regular signals. 
The upshot of the PnP approach is that many well-developed denoisers for natural images (e.g., the non-local means (NLM) \cite{buades2005non} and BM3D \cite{dabov2007image} denoisers) can be leveraged. In addition, as natural images are abundant, deep neural network-based denoisers can be easily trained \cite{zhang2018ffdnet,zhang2021plug}.

Nonetheless, applying PnP denoising to tackle the RME task is not straightforward.
Most existing PnP denoising-based methods operate directly in the data domain. Using PnP for RME in the data domain might incur nontrivial computations. This is because radio maps are high-order tensors, yet natural image denoisers are designed for grayscale or RGB images.
In addition, recoverability of tensor signals like radio maps under PnP denoising-based treatments is unclear---yet recoverability is a key consideration in such estimation problems.
Even for PnP algorithms' numerical behavior that is relatively well understood---convergence---existing analyses 
(e.g., those in \cite{chan2016plug,ryu2019plug,sreehari2016plug,Teodoro2019convergent}) mostly focused on formulations where the data is unconstrained. For radio maps that have various structural constraints, these analyses cannot directly apply.

\noindent
{\bf Contributions.} 
In this work, we propose a custom PnP approach for RME. Our detailed contributions are as follows:

\noindent
$\bullet$ {\bf A Latent Domain PnP Denoising Approach for RME}: We exploit the physical characteristics of radio maps to propose a tailored PnP algorithm for RME. We note that under reasonable conditions, radio maps admit a spatio-spectral decomposition, where the two latent components represent the spatial loss fields (SLFs) and the PSDs of emitters, respectively, in the region of interest \cite{Bazerque2011Splines,bazerque2009distributed,romero2017learning,Lee2017CGMcartography,Zhang2020SpectrumViaBlockTerm,Subash2024Quantized,Shrestha2022DeepSpectrum}.
Leveraging this decomposition,
we propose a PnP-based ADMM algorithm that only uses a grayscale image denoiser on the SLFs of the radio map.
This way, only one denoising operation is needed for each SLF per ADMM iteration.
The computational burden is substantially reduced compared to data-domain PnP (which needs many more denoising operations per ADMM iteration). 
In addition, explicitly using the decomposition structure of radio maps is naturally more robust to noise.

\noindent
$\bullet$ {\bf Recoverability Analysis}: We analyze
the recoverability of the ground-truth radio map under limited samples. By leveraging the connection between the implicit PnP regularization and a quadratic proximal operator of the linear denoisers \cite{sreehari2016plug,GAVASKAR2023109100,Gavaskar2021OnPnPLinear},
we establish sample complexity bounds for our latent-domain PnP framework under linear and symmetric image denoisers that are widely used in natural images \cite{sreehari2016plug,Teodoro2019convergent,chan2019PnPGraph,Gavaskar2021OnPnPLinear}.
As mentioned, the PnP denoising approach has been primarily treated as a computational framework, yet its statistical characterizations, such as recoverability, have been much less studied. An exception is \cite{liu2021recovery}, but the restricted isometric property (RIP) of the sensing matrix was needed there; and the result was for data-domain PnP denoising.
Our result fills this gap in the context of RME without using RIP.

\noindent
$\bullet$ {\bf Convergence Analysis}:
We also provide convergence support to the proposed ADMM algorithm with latent domain PnP denoising. 
Generalizing existing convergence results for PnP in the data domain \cite{chan2016plug},
our analysis shows that the latent domain approach still attains a fixed point convergence using a wide range of denoisers. This generalization is nontrivial, as the factorization of radio maps introduces nonconvex, bilinear terms in the objective function and the physical meaning of radio maps introduces non-negative constraints. These complex structures require more care to analyze. 
Beyond fixed point convergence, we also show that the algorithm converges to a \textit{Karush-Kuhn-Tucker} (KKT) point if linear denoisers are used. 

\medskip

We validate the effectiveness of the proposed PnP denoising-based RME approach over a diverse collection of datasets, including statistical model-based synthetic data, ray-tracing model-based synthetic data, and real-world data.

\vspace{2mm}

A conference version will appear at ICASSP 2025 
\cite{Xu2025RME}, presenting the basic idea of latent domain PnP denoising. This journal version includes an additional suite of denoisers, analyzes the recoverability of the formulated estimator, and also provides the convergence characterizations. In addition, unlike the conference version that only had limited experiments over a statistical radio map model, the journal version tests the algorithm over diverse datasets, including the statistical model \cite{goldsmith_2005}, the ray-tracing model \cite{sionna2022}, and two different datasets collected from the real world.

\medskip

\noindent \textbf{Notation:} In this paper, $x \in \mathbb{R}$, $\bm{x} \in \mathbb{R}^N$, $\bm{X} \in \mathbb{R}^{M\times N}$ and $\mathbm{X}\in \mathbb{R}^{M\times N \times K}$ denote a scalar, a vector, a matrix and a tensor, respectively. 
$\bm{X}(i,j)$, $\bm{X}(i,:)$ and $\bm{X}(:,j)$ denote the $(i,j)$th element, $i$th row and $j$th column of $\bm{X}$, respectively. Similar notations apply to vectors and tensors. ${\rm blkdiag}(\bm{X}_1,\ldots,\bm{X}_n)$ refers to a block-diagonal matrix with the $j$th diagonal block being $\bm{X}_j$. The range space of $\bm X$ is denoted as $\mathcal{R}(\bm X)$. The indicator function $\mathbb{1}[\bm{x}\in{\mathbb{X}}]$ equals $0$ if $\bm{x}\in{\mathbb{X}}$ and $+\infty$ otherwise. 
$\bm{1}_{\rm K} \in \mathbb{R}^K$ represents a vector with all elements being $1$. $[N]$ denotes the set of all integers from $1$ to $N$, i.e., $\{1,2,\ldots, N\}$. Symbols $\circ$ and $\oast$ denote the outer product and element-wise product, respectively.
The notation $[\bm X]_+$ means projecting $\bm X$ onto the nonnegative orthant.

\section{Problem Statement and Challenges Ahead}
\label{sec:ProblemandChallenges}
\subsection{Problem Statement}
\label{subsec:problemStatement}

\begin{figure}[t!]
\centering
    \includegraphics[width=0.8\linewidth]{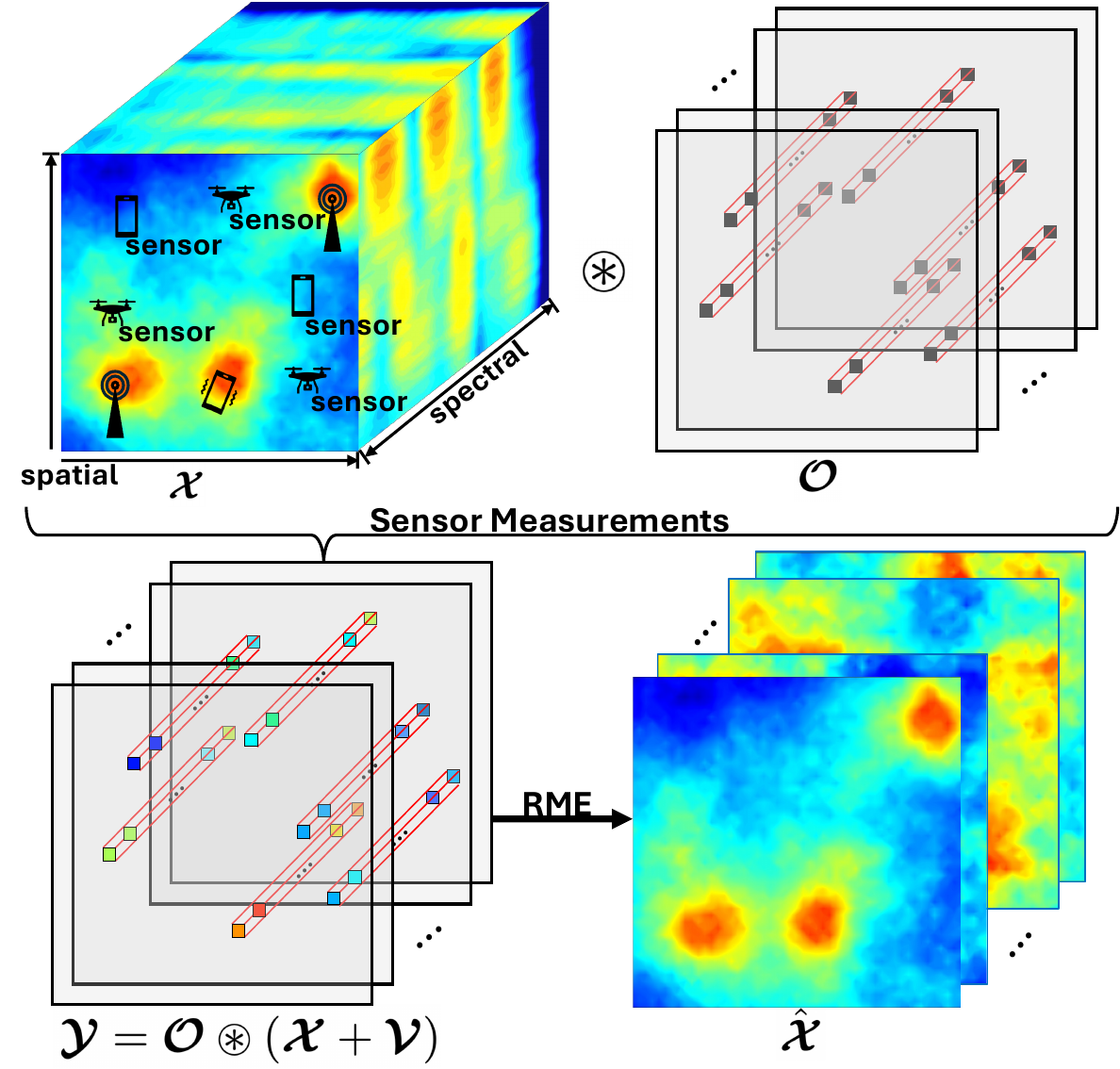}
    \caption{Illustration of the RME task. Upper: the spatio-spectral radio map (left) and the sampling tensor (right); bottom: use the sparse measurements (left) for estimation (right).}
    \label{fig:scproblem}
\end{figure}

Consider a scenario where $R$ emitters exist in a rectangular region. The emitters transmit signals over shared $K$ frequency bands. The signals interact with the environment (e.g., buildings and trees) and thus propagate irregularly---which results in a complex interference power propagation map over space and frequency; see Fig.~\ref{fig:scproblem}.
To be more specific, we adopt the problem setup in
\cite{Zhang2020SpectrumViaBlockTerm,sun2024integrated,Sun2022InterpolationMatrix,Bazerque2011Splines,Subash2024Quantized,Shrestha2022DeepSpectrum,teganya2021deep}, wherein the rectangular area is discretized into $M\times N$ grids. The frequency range is segmented into $K$ frequency bins. The goal is to estimate the discretized power spectrum density (PSD) in all the grids. These discretized PSDs can be {compactly} represented by a 3D radio map $\mathbm{X} \in \mathbb{R}^{M\times N \times K}$, in which $\mathbm{X}(m,n,k)$ denotes the PSD in grid $(m,n)$ and frequency $k$. Sensors are sparsely deployed over the grids, and their locations are collected by an index set
\begin{align}
   {\bm \varOmega\subseteq \{(m,n)| m\in [M], n \in [N]\}. \nonumber}
\end{align}
The notation $(m,n) \in \bm \varOmega$ indicates that there is a sensor in the grid whose coordinate is $(m,n)$.
We assume that the sensor at grid $(m,n)$ can measure the PSD of the received signal over all $K$ frequency bins. Therefore, the sampling process can be modeled by a binary tensor $\mathbm{O}\in \{0,1\}^{M\times N \times K}$, in which $\mathbm{O}(m,n,:) = \bm{1}_{\rm K}$ if $(m,n) \in \bm \varOmega$ and all other elements are zeros. Using these notations, the observed measurements can be represented as
\begin{align}\label{eqn:Yobserved}
    \mathbm{Y} = \mathbm{O}\oast (\mathbm{X} + \mathbm{V}),
\end{align}
where $\mathbm{V}\in \mathbb{R}^{M\times N \times K}$ stands for noise.
The RME task amounts to reconstructing $\mathbm{X}$ from $\mathbm{Y}$; also see Fig. \ref{fig:scproblem}. 

\subsection{Existing Approaches}
The RME task is clearly an under-determined, ill-posed inverse problem. Traditional inverse problem-solving techniques, such as kernel interpolation \cite{Bazerque2011Splines,kim2010cooperative,Boccolini2012Kriging}, low-rank matrix completion \cite{Sun2022InterpolationMatrix,chouvardas2016method}, and tensor completion \cite{Zhang2020SpectrumViaBlockTerm,Subash2024Quantized,sun2024integrated} have been applied to tackle RME.
These methods often work to a good extent, but the performance could deteriorate when the RF environment becomes more challenging, e.g., when heavy shadowing is present.
Recently, deep learning-based RME methods, e.g., \cite{Subash2024Quantized,Shrestha2022DeepSpectrum,Levie2021radiounet,teganya2021deep,romero2022radio,roger2023deep}, have shown promising {capability} to capture complex characteristics of radio maps.

{One common way to utilize deep learning for RME is training a neural network on a collected training set of radio maps $\{\mathbm{X}_{\ell}\}_{\ell=1}^L$.} The training set can be acquired using well-studied physical models \cite{goldsmith_2005}, numerical simulators \cite{sionna2022}, or historical data from the real world.
The training data $\mathbm{X}_{\ell}$ is expected to share similar characteristics (e.g., level of shadowing) of the target radio map of interest.
Using the training set, a neural predictor $\bm f_{\bm \theta}(\cdot):\mathbb{R}^{M\times N \times K} \rightarrow \mathbb{R}^{M\times N \times K}$ can be trained by
\begin{align}\label{eq:nntrain}
    \minimize_{\bm{\theta}} ~~\frac{1}{L} \sum_{\ell=1}^{L}   {\rm div} \left(\bm f_{\bm{\theta}}( \mathbm{O}_{\ell} \oast \mathbm{X}_{\ell}) ~||~ \mathbm{X}_{\ell}  \right),
\end{align}
where ${\rm div}(x||y)$ is a certain divergence (e.g., the Euclidean distance between $x$ and $y$).
The method in \eqref{eq:nntrain} conceptually summarizes the ideas in earlier learning-based works, e.g., \cite{teganya2021deep,Levie2021radiounet,romero2022radio,roger2023deep}.
The works in \cite{Shrestha2022DeepSpectrum,Subash2024Quantized} took a step further to consider training a generative model of $\mathbm{X}_{\ell}$, i.e.,
$ \mathbm{X}_\ell ={\bm g}_{\bm \beta}(\bm q_\ell)$,
where ${\bm g}_{\bm \beta}(\cdot):\mathbb{R}^d \rightarrow \mathbb{R}^{M\times N\times K}$ maps a latent low-dimensional representation to the data domain of radio maps.
Then, the inverse problem becomes finding the low-dimensional $\bm q$ from the observations. This generative model-based approach reduces the computational/modeling complexity of training and thus improves generalization \cite{Shrestha2022DeepSpectrum,Subash2024Quantized}.

Learning-based RME has clear advantages in handling complex RF environments. Nonetheless, their shortcomings are also obvious: First, if the training set has some distribution mismatches with the testing scenario, the performance often degrades substantially and even requires re-training. This hinders their capability of quickly adapting to new environments. Second, acquiring high-quality data is not trivial, considering the fast varying nature of RF configurations.

\subsection{Plug-and-Play Denoising}\label{subsec:pnpDenoise}

In recent years, the computational imaging community advocated a PnP denoising framework for inverse problem solving \cite{venkatakrishnan2013plug,Kamilov2023PNP,chan2016plug,sreehari2016plug,Teodoro2019convergent,wei2020tuning}.
The key idea of PnP is to incorporate well-developed image denoisers to assist handling tasks on other types of data (e.g., ocean SSF and MRI). 
For example, in a typical inverse problem of recovering $\bm{x}$ from measurement $\bm y = \bm A \bm x$, where $\bm A$ is a fat sensing matrix, applying the PnP framework starts from the following problem formulation:
\begin{subequations}\label{eq:pnp}
   \begin{align}
    &\minimize_{\bm x, \bm z}~\|\bm y-{\bm A}\bm x\|_2^2 + \lambda r(\bm z)\\
    &{\rm subject~to}:~\bm x = \bm z,
\end{align} 
\end{subequations}
where $r(\cdot)$ is an unspecified regularization that is supposed to capture characteristics of $\bm x$ and $\lambda$ is a regularization parameter. 
The interesting part of the PnP denoising framework is that $r(\cdot)$ is not explicitly specified, but will be realized by a denoising operation. To see this,
the ADMM \cite{Boyd2011ADMM} algorithm for solving \eqref{eq:pnp} consists of the following steps in the $j$th iteration:
\begin{subequations}\label{eq:admmupdate}
    \begin{align}
        \bm x^{(j+1)} &\leftarrow \argmin_{\bm{x}} \|\bm y-{\bm A}\bm x\|_2^2 +\frac{\rho}{2}\|\bm x-\tilde{\bm x}^{(j)} \|_2^2\\
        \bm z^{(j+1)} &\leftarrow \argmin_{\bm{z}} \frac{\rho}{2\lambda}\|\bm z - \tilde{\bm z}^{(j)}\|_2^2 + r(\bm z) \label{eq:denoise}\\
        \bm u^{(j+1)} &\leftarrow \bm u^{(j)} + \bm x^{(j+1)} - \bm z^{(j+1)},\label{eq:DualUpdate}
    \end{align}
\end{subequations}
where $\tilde{\bm x}^{(j)} = \bm z^{(j)} - \bm u^{(j)}$, $\tilde{\bm z}^{(j)} = \bm x^{(j+1)} +\bm u^{(j)}$, $\bm u^{(j)}$ is the dual variable, and $\rho$ is the augmented Lagrangian parameter. Notably, the $\bm z$-update step \eqref{eq:denoise} can be viewed as a {\it denoising} process---that is, one hopes to use $\bm z^{(j+1)}$ to extract the signal from the noisy version $\tilde{\bm z}^{(j)}$, with the structural regularization $r(\bm z)$ acting as the prior information.
Based on this observation, it is well-motivated to replace the step in \eqref{eq:denoise} by the following \cite{venkatakrishnan2013plug,Kamilov2023PNP}:
\begin{align}
    \bm z^{(j+1)} \leftarrow \bm{D}_{\sigma}(\tilde{\bm z}^{(j)}),
    \label{eqn:deepDenoiser}
\end{align}
where $\bm{D}_{\sigma}$ denotes the denoiser operating at the noise level with a $\sigma^2$ variance. 

The major postulate of the PnP framework is that
many different types of data (e.g., natural images \cite{Gavaskar2021OnPnPLinear,chan2016plug,zhang2021plug}, MRI images \cite{wei2020tuning,liu2020rare}, ocean SSFs \cite{li2024zero}, and tomographic images \cite{Majee2021CT,sreehari2016plug}) share similar denoising processes. As a consequence, one can use well-developed natural image denoisers for many other types of signals, sparing developing or training new denoisers for them. Some widely used denoisers are as follows:

\subsubsection{Linear Denoisers} \label{subsubsec:linearDenoiser}
Many denoisers from image processing admit the following form:
\begin{align}
    \bm{D}_{\sigma} (\tilde{\bm{z}}^{(j)}) = \bm{W} \tilde{\bm{z}}^{(j)},
    \label{eqn:NLMdenoiser}
\end{align}
where $\bm{W} \in \mathbb{R}^{MN\times MN}$ is the denoising matrix, which is commonly constructed to be symmetric \cite{Milanfar2013tour}. 
Classical denoisers, e.g., box filters \cite{szeliski2022computer}, Gaussian filters \cite{szeliski2022computer}, and Gaussian mixture model (GMM) denoisers \cite{yu2011solving,Teodoro2019convergent} are all linear denoisers.
In addition, many kernel denoisers follow a similar form of \eqref{eqn:NLMdenoiser}, but compute the filter $\bm W(\tilde{\bm{z}}^{(j)})$ as a function of $\tilde{\bm{z}}^{(j)}$; see the \textit{non-local means} (NLM) denoiser \cite{milanfar2024denoising} and its symmetrized modifications, e.g., doubly stochastic gradient NLM (DSG-NLM) denoiser \cite{sreehari2016plug}.
Nonetheless, 
in ADMM PnP denoising, it is common practice to stop updating $\bm{W}(\tilde{\bm z}^{(j)})$ after a certain number of iterations \cite{sreehari2016plug,Gavaskar2021OnPnPLinear,GAVASKAR2023109100,chan2019PnPGraph}, e.g., fixing $\bm W=\bm{W}(\tilde{\bm z}^{(10)})$ after the $10$th iteration. Hence, such denoisers are essentially linear after several iterations.

\subsubsection{Nonlinear Denoisers}
Non-linear denoisers are with more complex designs and could be more powerful than linear denoisers. However, they are more challenging to analyze. The most commonly adopted non-linear denoisers include BM3D and those based on deep neural networks.

\noindent
$\bullet$ {\bf BM3D Denoiser.} The BM3D denoiser \cite{dabov2007image}  computes the data-dependent filter $\bm{W}(\tilde{\bm z}^{(j)})$ in a complex and structured manner. Specifically, BM3D employs a two-stage denoising process, comprising hard-thresholding followed by Wiener filtering. Both stages operate on 3D blocks constructed by stacking similar patches from the image.  
In the field of image denoising, BM3D is widely used as a benchmark \cite{Milanfar2013tour}.

\noindent
$\bullet$ {\bf Deep Neural Denoiser.} Deep denoisers, e.g., \cite{zhang2018ffdnet,zhang2021plug}, can be learned via training a neural network $\bm D_\sigma(\cdot):\mathbb{R}^{MN}\rightarrow \mathbb{R}^{MN}$:
\begin{align}
    \minimize_{\bm D_{\sigma}(\cdot)}~\frac{1}{N}\sum_{\ell=1}^N\| \bm x_\ell - \bm D_{\sigma}(\bm x_\ell +\bm v_\ell)  \|_2^2,
\end{align}
where $\{\bm x_\ell\}_{\ell=1}^N$ is the training set, and $\bm v_\ell$ is random noise with variance $\sigma^2$. As natural images are abundant, training such denoisers is considered a relatively easy task.

\subsection{Challenges In The Context of RME}
\label{subsec:challenges}
Directly applying the PnP framework \eqref{eq:pnp} and \eqref{eq:admmupdate} to the RME task leads to the following problem formulation:
\begin{align}\label{eqn:RMEproblem_data}
    \minimize_{\mathbm{X}, \mathbm{\bm Z} }&~ \| \mathbm{O} \oast (\mathbm{Y} - \mathbm{X}) \|_{\rm F}^2 +  \lambda r(\mathbm{Z}), \nonumber\\
    {\rm subject~to}:&~\mathbm{X} = \mathbm{\bm Z} 
\end{align}
and a denoiser applied onto $\mathbm{X} + \mathbm{U}$ is needed for implementing the step in \eqref{eq:denoise}. 
However, $\mathbm{X}$ is a 3D tensor of size $M \times N \times K$, but typical image denoisers are trained on 2D grayscale or RGB images. 
To accommodate this mismatch, a way is to apply the grayscale image  denoiser repeatedly for $K$ times in the step \eqref{eq:denoise}, resulting in high computational complexity (see, e.g., \cite{li2024zero,Liu2022HSIpnp}).
Another challenge lies in theoretical understanding. Statistical characteristics of the PnP denoising approach, e.g., sample complexity and recoverability, have received limited attention. {Existing studies analyzing the recoverability of the basic inverse model \eqref{eq:pnp} rely on RIP assumptions of the sensing matrix $\bm{A}$ \cite{liu2021recovery,GAVASKAR2023109100}.}
However, these do not answer critical questions in the context of RME---e.g., how many sensor measurements are sufficient to recover the ground-truth radio map to a certain accuracy? In addition, 
many works analyzed the convergence properties of PnP ADMM when $\mathbm{X}$ is not constrained other than having $r(\cdot)$ as regularization; see, e.g., \cite{chan2016plug,ryu2019plug,sreehari2016plug,Teodoro2019convergent}.
Nonetheless, when the radio map $\mathbm{X}$ has complex structural constraints---as we will exploit---designing the ADMM updates to accommodate these constraints while retaining convergence guarantees is not straightforward.

\section{Proposed method}
\label{sec:method}
\subsection{Problem Formulation}
\begin{figure}
\centering
    \centering
        \includegraphics[width=0.8\linewidth]{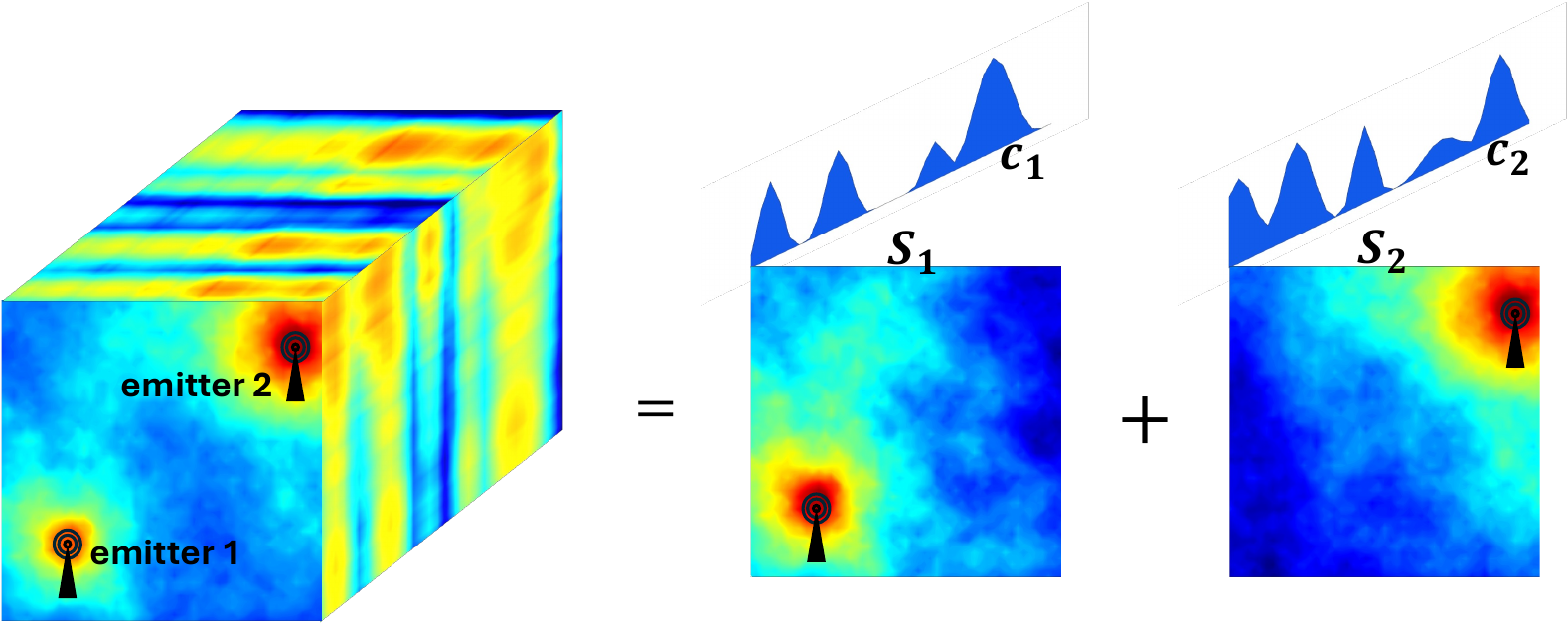}
    \caption{Illustration of the decomposition model of radio maps.}
    \label{fig:X=SC}
\end{figure}

To propose a PnP denoising method that is tailored for RME, our idea is to apply the denoisers onto latent factors of the radio maps. To be more specific, we use the following model for radio maps:
\begin{align}
    \mathbm{X}(m,n,k) = \sum_{r=1}^R\bm{S}_r(m,n) \bm{c}_r(k) \Leftrightarrow \mathbm{X} = \sum_{r=1}^R \bm{S}_r \circ \bm{c}_r, 
    \label{eqn:X=SC}
\end{align}
where $\bm{S}_r \in \mathbb{R}^{M\times N}$ represents the \textit{spatial loss field} (SLF) of emitter $r$ over the area, $\bm{c}_r \in \mathbb{R}^K$ represents emitter $r$'s PSD across the $K$ frequency bands, and $\circ$ denotes the outer product, i.e., $[\bm X\circ \bm y]_{i,j,k}=\bm X(i,j)\bm y(k)$. Both $\bm{S}_r$ and $\bm{c}_r$ are {non-negative} according to their physical interpretations.
An illustration is shown in Fig. \ref{fig:X=SC}. 
This model has been proven effective for radio maps that are measured over a relatively narrow band \cite{Bazerque2011Splines,sun2024integrated,romero2017learning,teganya2021deep,Shrestha2022DeepSpectrum,Subash2024Quantized}.

Under \eqref{eqn:X=SC}, we propose the following formulation:
\begin{subequations}\label{eq:ourformulation}
\begin{align}
    & \minimize_{\{\bm{S}_r, \bm{c}_r \}_{r=1}^R} \big\| \mathbm{O} \oast \big(\mathbm{Y} -  \sum_{r=1}^R \bm{S}_r \circ \bm{c}_r \big) \big\|_{\rm F}^2 \nonumber\\
    & \qquad \qquad \qquad+ \lambda \sum_{r=1}^R r(\bm{S}_r) + \zeta \sum_{r=1}^R \bm{c}_r^\top\bm{c}_r,
    \label{eqn:problem_withrank} \\
    &{\rm subject~to}:~\bm S_r\geq \bm 0,~\bm c_r\geq \bm 0.~\forall r\in[R],
\end{align}
\end{subequations}
where $\mathbm{Y}$ was defined in \eqref{eqn:Yobserved}, $r(\cdot)$ is the denoiser related regularization, the term $\bm{c}_r^\top\bm{c}_r$ is adopted to make the formulation more stable, and $\lambda$ and $\zeta$ are regularization parameters.

In \eqref{eq:ourformulation}, the term $r(\cdot)$ is applied onto the SLFs, which constitute the spatial latent factor in \eqref{eqn:X=SC}. As one will see, this assists in designing a more efficient PnP denoising algorithm than that derived from the data-domain formulation in \eqref{eqn:RMEproblem_data}.

\subsection{Proposed Algorithm}
\label{subsec:admm}
We propose to solve \eqref{eq:ourformulation} via ADMM. First, Eq.~\eqref{eq:ourformulation} is recast into the following:
\begin{align}
    & \minimize_{\{\bm{S}_r, \bm Z_r, \bm{c}_r\}_{r=1}^R} \bigg\| \mathbm{O} \oast \bigg(\mathbm{Y} -  \sum_{r=1}^R \bm{S}_r \circ {\bm{c}}_r \bigg) \bigg\|_{\rm F}^2  \nonumber \\
    & \qquad \qquad \qquad \qquad + \lambda \sum_{r=1}^R r(\bm{Z}_r) + \zeta \sum_{r=1}^R \bm{c}_r^\top\bm{c}_r, \nonumber \\
    & {\rm subject~to}:~\bm{S}_r\geq \bm{0},~\bm{S}_r = \bm{Z}_r,~\bm{c}_r \geq \bm{0},~\forall r \in[R].
    \label{eqn:admmAuxiliary}
\end{align}
The augmented Lagrangian of \eqref{eqn:admmAuxiliary} is
\begin{align}
    &L_\rho(\{\bm{S}_r,\bm{c}_r,\bm{Z}_r,\bm{\Psi}_r\}_{r=1}^R) = \bigg\| \mathbm{O} \oast \bigg(\mathbm{Y} -  \sum_{r=1}^R \bm{S}_r \circ {\bm{c}}_r \bigg) \bigg\|_{\rm F}^2  \nonumber \\
    & +  \lambda \sum_{r=1}^R r(\bm{Z}_r) + \sum_{r=1}^R \bigg(\frac{\rho}{2} \| \bm{S}_r - \bm{Z}_r + \bm{\Psi}_r\|_{\rm F}^2 -\frac{\rho}{2}\| \bm{\Psi}_r \|_{\rm F}^2 \bigg) \nonumber \\
    & + \zeta \sum_{r=1}^R \bm{c}_r^\top\bm{c}_r,
    \label{eqn:Lagaragian}
\end{align}
where $\rho$ is the augmented Lagrangian parameter, and $\bm{\Psi}_r \in \mathbb{R}^{M\times N}$ is the {scaled} dual variable \cite{Boyd2011ADMM}. 

The ADMM algorithm iteratively updates the primal variables $\{\bm{Z}_r\}_{r=1}^R$, $\{\bm{S}_r,\bm{c}_r\}_{r=1}^R$, and dual variables $\{\bm{\Psi}_r\}_{r=1}^R$. The detailed updates are as follows.

\subsubsection{The \texorpdfstring{$\{\bm{Z}_r\}_{r=1}^R$}{Z}-update}
Minimizing \eqref{eqn:Lagaragian} w.r.t. $\{\bm{Z}_r\}_{r=1}^R$ can be decoupled for each $r$; that is,
\begin{align}
    \bm Z_r\leftarrow \arg\min_{\bm{Z}_r}  r(\bm{Z}_r) + \frac{\rho}{2\lambda}\| \bm{S}_r - \bm{Z}_r + \bm{\Psi}_r\|_{\rm F}^2.
    \label{eqn:prox_z}
\end{align}
Similar to solving \eqref{eq:denoise}, solving \eqref{eqn:prox_z} w.r.t. each $\bm{Z}_r$ can be seen as a denoising problem with noise variance $\lambda/\rho$. Consequently, $\bm{Z}_r$ can be updated through
\begin{align}
    \bm{Z}_r = \bm{D}_{\sqrt{{\lambda}/{\rho}}} (\bm{S}_r+\bm{\Psi}_r),
    \label{eqn:z_update}
\end{align}
where we set $\sigma = \sqrt{\lambda/\rho}$ following the literature \cite{venkatakrishnan2013plug,Kamilov2023PNP}.
From \eqref{eqn:prox_z} and \eqref{eqn:z_update}, it is clear that the denoiser only needs to be applied for $R$ times, which is normally much less than $K$. This saves much time compared to solving the denoising step under the data-domain formulation in \eqref{eqn:RMEproblem_data}.

\subsubsection{The \texorpdfstring{$\{\bm{S}_r,\bm{c}_r\}_{r=1}^R$}{S,c}-update}
\begin{figure}[!tb]
    \centering
        \includegraphics[width=0.55\linewidth]{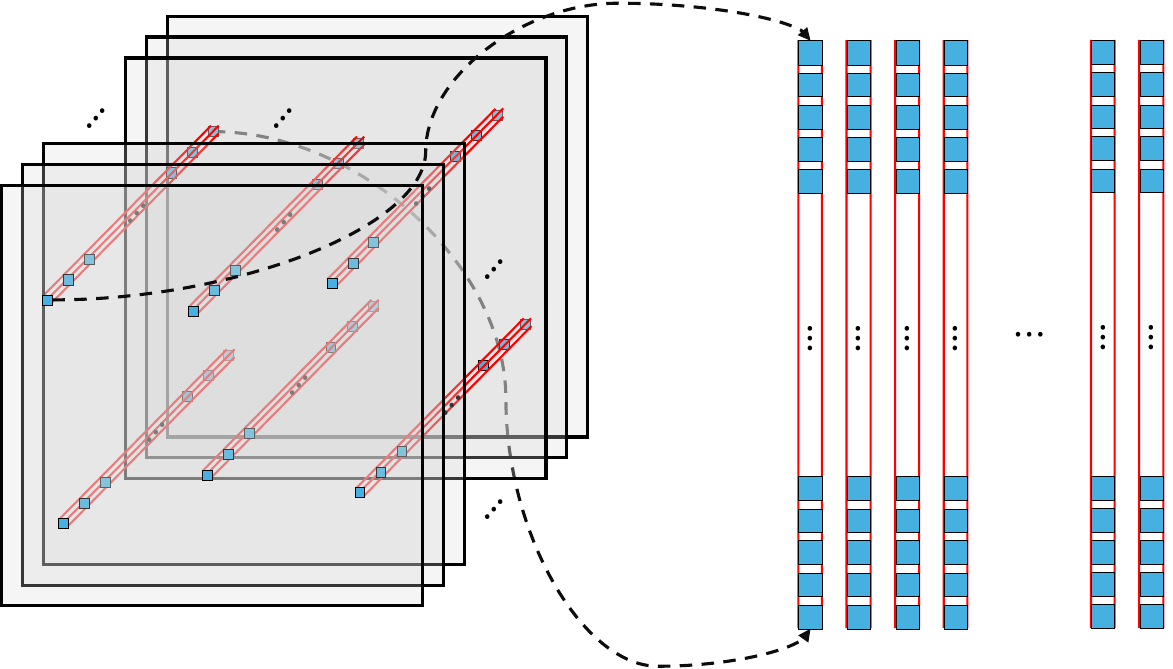}
    \caption{Illustration of matricization of the measurements. Left: the measurements $\mathbm{O}\oast \mathbm{Y}$; right: $\bm{Y}(:,\bm{\varOmega}_{\text{vec}})$.}
    \label{fig:sampling_mode3}
\end{figure}

We first re-express all terms in \eqref{eqn:Lagaragian} into vector or matrix forms. We introduce the matricization of $\mathbm{Y}$, i.e.,
\begin{align}
    \bm{Y} = [\mathbm{Y}(1,1,:),\mathbm{Y}(2,1,:),\ldots, \mathbm{Y}(M,N,:)].
\end{align}
In addition, we define the set $\OmegaVec \subseteq [MN]$ such that $$(m,n) \in \bm \varOmega \Leftrightarrow n(M-1)+m \in \bm \varOmega_{\text{vec}}.$$ Then, the non-zero elements in $\mathbm{O} \oast \mathbm{Y}$ can be represented with $\bm{Y}(:,\OmegaVec)$, which takes out all the measured columns in $\bm{Y}$. 
An illustration of $\bm{Y}(:,\bm{\varOmega}_{\text{vec}})$ is provided in Fig. \ref{fig:sampling_mode3}. 
The following equation holds if there is no noise:
\begin{align}
    \mathbm{Y}(:,\OmegaVec) = \sum_{r=1}^R \bm{c}_r \bm{s}_r(\OmegaVec)^\top,
\end{align}
where $\bm{s}_r={\rm vec}(\bm{S}_r)$. By further defining $\OmegaC = [MN]\backslash \OmegaVec$, the $\{\bm{S}_r,\bm{c}_r\}_{r=1}^R$-subproblem can be rewritten as follows:
\begin{align}
    &\minimize_{\{\bm{s}_r,\bm{c}_r\}_{r=1}^R}  \Big\| \bm{Y}(:,{\bm \varOmega}_{\text{vec}})  -  \sum_{ r }^R {\bm{c}}_{r} {\bm{s}_r(\OmegaVec)^\top} \Big\|_{\rm F}^2 \nonumber \\
     &+ \frac{\rho}{2} \sum_{r=1}^R\| \bm{s}_r(\OmegaVec) - \bm{z}_r(\OmegaVec) + \bm{\psi}_r(\OmegaVec)\|_{\rm F}^2 + \zeta \sum_{r=1}^R \bm{c}_r^\top\bm{c}_r\nonumber \\
     &+ \frac{\rho}{2} \sum_{r=1}^R\| \bm{s}_r(\OmegaC) - \bm{z}_r(\OmegaC) + \bm{\psi}_r(\OmegaC)\|_{\rm F}^2, \label{eqn:admmPrimal} \\
     & {\rm subject~to}:~\bm{s}_r \geq \bm 0,~\bm{c}_r\geq \bm 0,~\forall r\in[R], \nonumber    
\end{align}
in which $\bm{z}_r={\rm vec}(\bm Z_r)$ and $\bm{\psi}_r={\rm vec}(\bm \Psi_r)$. It is obvious that $\{\bm{s}_r(\OmegaVec),\bm{c}_r\}_{r=1}^R$ and $\{\bm{s}_r(\OmegaC)\}_{r=1}^R$ are uncoupled in \eqref{eqn:admmPrimal}, and thus can be found separately.

Minimization w.r.t. $\{\bm{s}_r(\OmegaVec),\bm{c}_r\}_{r=1}^R$ can be seen as a \textit{non-negative matrix factorization} (NMF) problem with regularization. There exist many off-the-shelf algorithms for NMF. Here, we adopt the \textit{hierarchical alternative least square} (HALS) method \cite{zhou2014nonnegative}, which updates $\bm{s}_r(\OmegaVec)$ and $\bm{c}_r$ for different $r$ iteratively in a block coordinate descent (BCD) manner. The detailed updates are as follows:
\begin{align}
    &\bm{s}_r(\OmegaVec) = \bigg[ \frac{1}{{\bm{c}}_r^\top{\bm{c}}_r + \frac{\rho}{2}} \bigg( \frac{\rho}{2}\Big(\bm{z}_r(\OmegaVec)-\bm{\psi}_r(\OmegaVec)\Big)\nonumber \\
    &  + \Big( \bm{Y}(:,\OmegaVec) - \sum_{r'\neq r}^R {\bm{c}}_{r'} \bm{s}_{r'}(\OmegaVec)^\top\Big)^\top{\bm{c}}_r \bigg) \bigg]_+.
    \label{eqn:x_update_observed_basic}
\end{align}
\begin{align}
    \bm{c}_r = \left[ \frac{\Big( \bm{Y}(:,\OmegaVec) - \sum_{r'\neq r}^R {\bm{c}}_{r'}{\bm{s}_{r'}(\OmegaVec)}^\top\Big)\bm{s}_r(\OmegaVec)}{{\bm{s}_r(\OmegaVec)}^\top\bm{s}_r(\OmegaVec) + \zeta } \right]_+.
    \label{eqn:c_update}
\end{align}

For $\bm{s}_r(\OmegaC)$, it can be solved by minimizing the last line in \eqref{eqn:admmPrimal}, leading to the update
\begin{align}
    \bm{s}_r(\OmegaC) = [\bm{z}_r(\OmegaC) - \bm{\psi}_r(\OmegaC)]_+.
    \label{eqn:x_update_unobserved_basic}
\end{align}

\subsubsection{The \(\{\mathbf{\Psi}_{r}\}_{r=1}^R\)-update}
For the dual variables $\bm{\Psi}_r$, they can be updated following the standard ADMM scheme \eqref{eq:DualUpdate} as
\begin{align}
    \bm{\Psi}_r = \bm{\Psi}_r + \bm{S}_r - \bm{Z}_r.
    \label{eqn:dualupdate}
\end{align}

\medskip
The proposed {\it latent domain ADMM plug-and-play denoising} (\texttt{LaPnP}) algorithm is summarized in Algorithm \ref{alg:admm_pnp}.
\begin{algorithm}[tb!]
\caption{\texttt{LaPnP} for solving \eqref{eq:ourformulation}.}
\SetAlgoLined
  \textbf{Input:} the observed tensor ${\mathbm{Y}}$, sampling tensor $\mathbm{O}$, number of emitters $R$, augmented Lagrangian parameter $\rho$, $\eta\in(0,1]$ and $\gamma > 1$\;

  Initialize $\{\bm{s}_r, \bm{c}_r, \bm{z}_r, \bm{\psi}_r\}_{r=1}^R$\;
 
 \While{Not Converged}{
    Update $\{\bm{Z}_r\}_{r=1}^R$ with \eqref{eqn:z_update}\;
    \For{$j=1$ to $J$}{
        Update $\{\bm{s}_r,\bm{c}_r\}_{r=1}^R$ with \eqref{eqn:x_update_observed_basic}-\eqref{eqn:x_update_unobserved_basic}\;
    }
    Update $\{\bm{\psi}_r\}_{r=1}^R$ with \eqref{eqn:dualupdate}\;
    \If{$\triangle_t \geq \eta \triangle_{t-1}$}
        {
            $\rho \gets \gamma \rho$\;
        }
}
 \label{alg:admm_pnp}
\end{algorithm}

\subsection{More Discussions}
\label{subsec:init}
\subsubsection{Initialization}
Eq.~\eqref{eq:ourformulation} is a non-convex problem, and thus good initialization often assists in attaining high-quality solutions. Here we employ a two-step approach as in \cite{Shrestha2022DeepSpectrum} to initialize $\{\bm{s}_r,\bm{c}_r\}_{r=1}^R$. First of all, $\{\bm{s}_r(\OmegaVec),\bm{c}_r\}_{r=1}^R$ is initialized by solving the following problem
\begin{align}
    & \minimize_{\{\bm{s}_r(\OmegaVec), \bm{c}_r\}_{r=1}^R} \bigg\|  \bm{Y}(:,\OmegaVec) -  \sum_{r=1}^R \bm{c}_r \bm{s}_r(\OmegaVec)^\top \bigg\|_{\rm F}^2,\nonumber\\
     &{\rm subject~to}:~\bm{s}_r(\OmegaVec)\geq \bm 0,~\bm{c}_r \geq \bm 0,~\forall r\in[R],
    \label{eqn:nmf}
\end{align}
which is a standard NMF problem. The successive projection algorithm (SPA) \cite{Gillis2014SPA,fu2015self}---a self-dictionary algorithm employing a greedy search scheme---is adopted to solve \eqref{eqn:nmf}. 
This initialization is particularly useful when the PSDs are relatively sparse. After that, nearest neighbors interpolation method is adopted to initialize the unmeasured elements $\{\bm{s}_r(\OmegaC)\}_{r=1}^R$.
For $\{\bm{z}_r,\bm{\psi}_r\}_{r=1}^R$, they are initialized as all zeros.

\subsubsection{Inexact HALS} Note that the HALS algorithm for solving the $\{\bm S_r,\bm c_r\}_{r=1}^R$-subproblem is an iterative algorithm. It could take a long time if one runs HALS until convergence is reached---which makes Algorithm~\ref{alg:admm_pnp} have an undesired slow inner loop. In this work, we use an inexact updating strategy. That is, we fix the number of iterations of HALS to be $J$, where $J$ is a small integer (e.g., we set $J=20$ in our experiments). As $J$ is small, the $\{\bm S_r,\bm c_r\}_{r=1}^R$-subproblem might not have been solved exactly.
Nonetheless, as one will see, even under such an inexact subproblem solving scheme, the ADMM algorithm is still ensured to converge.

\subsubsection{The $\rho$-update} 
Increasing $\rho$ during the ADMM iterations enhances the convergence performance \cite{chan2016plug,goldstein2014fast}. 
Following the strategy in \cite{chan2016plug}, we update $\rho$ by observing the following residual:
\begin{align}
    \triangle_{t} = \frac{1}{\sqrt{MN}}\sum_{r=1}^R & ( \|\bm{s}_r^{(t)} - \bm{s}_r^{(t-1)} \|_2 + \|\bm{z}_r^{(t)} - \bm{z}_r^{(t-1)} \|_2 \nonumber\\
    & + \|\bm{\psi}_r^{(t)} - \bm{\psi}_r^{(t-1)} \|_2 ),
    \label{eq:triangle}
\end{align}
where $t$ is the iteration index.
Using pre-specified $\eta \in (0,1]$ and $\gamma > 1$, $\rho$ is replaced with $\gamma \rho$ if $\triangle_{t} \geq \eta \triangle_{t-1}$; otherwise, it remains unchanged. We set $\eta = 0.95$ and $\gamma = 1.1$.

\section{Performance Analysis}
In this section, we provide performance characterizations of the proposed method.
\subsection{Recoverability Analysis under Linear Denoising}
\label{subsec:recoverability}
The first question lies in the recoverability of our formulation \eqref{eq:ourformulation}---that is, denoting $\{\bm S_r^\star,\bm c_r^\star\}_{r=1}^R$ as any optimal solution of \eqref{eq:ourformulation} and $\mathbm{X}^\natural$ as the ground-truth radio map, can we establish $\mathbm{X}^\star \approx \mathbm{X}^\natural$, where $\mathbm{X}^\star = \sum_{r=1}^R \bm S_r^\star \circ\bm c_r^\star$?

If $r(\cdot)$ has an explicit expression, then recoverability analysis is similar to that of structural tensor/matrix completion; see, e.g., \cite{Shrestha2022DeepSpectrum,Subash2024Quantized}.
However, in our case, the structural regularization $r(\cdot)$ is {\it implicitly} realized by the denoiser $\bm D_{\sigma}(\cdot)$. 
Analyzing recoverability under unknown $r(\cdot)$ in the latent domain of $\mathbm{X}^\natural$ is in general hard. The closest effort was in \cite{liu2021recovery} under a compressive sensing setting. However, it works in the data domain and relies on the RIP condition of the sensing matrix---both do not apply to our setting.
Establishing recoverability under the PnP framework with complex denoisers $\bm D_{\sigma}$, e.g., deep denoisers or kernel denoisers, poses a challenging task. In this work, we provide recoverability analysis of \eqref{eq:ourformulation} for linear denoisers. To proceed, we make the following assumption:
\begin{assumption}\label{assump_linearDenoiser}
    For $\bm{D}_{\sigma}(\bm x) = \bm W \bm x$, $\bm W \in \mathbb{R}^{MN\times MN}$ is non-negative, symmetric, and irreducible \cite{horn2012matrix}. In addition, the eigenvalues of $\bm W$, denoted as $\lambda_i(\bm W)$ for $i=1,\ldots, MN$, are arranged in descending order with $\lambda_1(\bm W) \leq 1$ and $\lambda_i(\bm W) \in [0,1)$ for all $i>1$.
\end{assumption}
Many linear denoisers (e.g., those mentioned in Sec. \ref{subsec:pnpDenoise}) are non-negative, symmetric, and irreducible \cite{milanfar2024denoising,Milanfar2013tour,Teodoro2019convergent,sreehari2016plug}. 
In addition, denoisers such as DSG-NLM, the box denoiser, the Gaussian filters, and the GMM denoiser (see Sec \ref{subsec:pnpDenoise}) are all known to satisfy $\lambda_1(\bm W) \leq 1$ and $\lambda_i(\bm W) \in [0,1)$ for all $i>1$ \cite{Milanfar2013tour,Teodoro2019convergent}.
For example,
the work~\cite{sreehari2016plug} showed that the denoising matrix $\bm W$ in DSG-NLM is doubly stochastic (i.e., both row- and column-stochastic) and primitive \cite{berman1994nonnegative}.  
Therefore, by the Perron-Frobenius theorem \cite{berman1994nonnegative}, such $\bm{W}$'s largest eigenvalue is $1$ and the algebraic multiplicity of this eigenvalue is 1 \cite{sreehari2016plug}. 

Using Assumption~\ref{assump_linearDenoiser},
we invoke the following classical result from image denoising \cite{sreehari2016plug,Teodoro2019convergent,chan2019PnPGraph,Gavaskar2021OnPnPLinear,GAVASKAR2023109100}:
\begin{lemma}\label{lem:proximal}
Consider a proximal operator:
\begin{align}
   \bm{z}^\star = \arg\min_{\bm z}   \frac{\rho}{2\lambda} \|  \bm E - \bm Z  \|_{\rm F}^2 + r(\bm Z),
\end{align}
where $\bm Z={\rm mat}(\bm z)$.
If $\bm z^\star = \bm W \bm e$ with $\bm e={\rm vec}(\bm E)$, and
$\bm W$ satisfies Assumption \ref{assump_linearDenoiser}, then the corresponding $r(\cdot)$ can be expressed as
\begin{align}
    r(\bm{Z}) = \frac{\rho}{2\lambda} \bm{z}^\top \tilde{\bm{Q}} (\tilde{\bm{\Lambda}}^{-1} - \bm{I} ) \tilde{\bm{Q}}^\top  \bm{z}  + \mathbb{1}[\bm{z}\in{\mathcal{R}({\tilde{\bm{Q}}})}],
    \label{eqn:regularizaitonLinear}
\end{align}
where $\bm W=[ \tilde{\bm Q},\tilde{\bm Q}^c] {\rm bkdiag}(\tilde{\bm \Lambda},\bm 0) [ \tilde{\bm Q},\tilde{\bm Q}^c]^\top  $ represents the eigen-decomposition of $\bm W$, and $\tilde{\bm Q}\in\mathbb{R}^{MN\times L}$ and $\tilde{\bm Q}^c$ are the eigenvectors associated with the nonzero eigenvalues in $\tilde{\bm \Lambda}\in\mathbb{R}^{L\times L}$ and the zero eigenvalues in $\bm 0$, respectively.
\end{lemma}

With Lemma \ref{lem:proximal}, Problem \eqref{eq:ourformulation} can be expressed in the following explicit format:
\begin{align} \label{eqn:ourformulationExplicit}
    & \minimize_{\{\bm{S}_r, \bm{c}_r \}_{r=1}^R} \bigg\| \mathbm{O} \oast \bigg(\mathbm{Y} -  \sum_{r=1}^R \bm{S}_r \circ {\bm{c}}_r \bigg) \bigg\|_{\rm F}^2 \nonumber \\
     & \quad + \frac{\rho}{2}  \sum_{r=1}^R \bm{s}_r^\top\tilde{\bm{Q}}_r (\tilde{\bm{\Lambda}}_r^{-1} - \bm{I} ) \tilde{\bm{Q}}_r^\top \bm{s}_r + \zeta \sum_{r=1}^R \bm{c}_r^\top\bm{c}_r, \\
    &{\rm subject~to}:~\bm S_r\geq \bm 0,~\bm{s}_r \in \mathcal{R}(\tilde{\bm Q}_r),~\bm c_r\geq \bm 0, ~\forall r\in[R],\nonumber
\end{align}
where the constraint $\bm{s}_r \in \mathcal{R}(\tilde{\bm Q}_r)$ comes from the indicator function in \eqref{eqn:regularizaitonLinear}. We will denote the target function in \eqref{eqn:ourformulationExplicit} as $v_{\rm obj}(\{\bm{S}_r,\bm{c}_r\}_{r=1}^R)$.
The optimal solution to \eqref{eqn:ourformulationExplicit} can be characterized with the following lemma:

\begin{lemma}\label{lem:boundedset}
Assume $\bm W_r=[ \tilde{\bm Q}_r,\tilde{\bm Q}_r^c] {\rm bkdiag}(\tilde{\bm \Lambda}_r,\bm 0) [\tilde{\bm Q}_r,\tilde{\bm Q}_r^c]^\top  $ satisfies Assumption \ref{assump_linearDenoiser} for $r\in[R]$, and $|\bm \varOmega|>1$. 
Denote $(\{\bm S_r^\star,\bm c_r^\star\}_{r=1}^R)$ as any optimal solution of \eqref{eq:ourformulation}. Then, $\bm G= \bm{Q}^\top{\bm \Xi}^\top{\bm \Xi} \bm{Q} + \frac{\rho}{2}  (\bm{\Lambda}^{-1} - \bm{I}) $ has full rank, with
\begin{align*}
    \bm Q &={\rm blkdiag}(\tilde{\bm Q}_1,\ldots,\tilde{\bm Q}_R),~
    \bm \Lambda = {\rm blkdiag}(\tilde{\bm \Lambda}_1,\ldots,\tilde{\bm \Lambda}_R),\\
    \bm \Xi &= [ {\rm diag}(\bm{o})\otimes \bm{c}_1^\star, {\rm diag}(\bm{o})\otimes \bm{c}_2^\star, \ldots, {\rm diag}(\bm{o})\otimes \bm{c}_R^\star],
\end{align*}
where $\bm o\in \{0,1\}^{MN}$ with $\bm o(\OmegaVec)=\bm 1$ and $\bm o(\OmegaC)= \bm 0$.
In addition, we have
\begin{subequations}\label{eqn:boundedSolution}
    \begin{align}
    \sum_{r=1}^R\| \bm{c}_r^\star \|_2^2 & \leq \frac{{v_{\rm obj}^\natural}}{\zeta} \triangleq {\alpha},
    \label{eqn:cbound}\\
    \sum_{r=1}^R\|\bm{S}_r^\star\|_{\rm F}^2 & \leq \frac{\left(\sqrt{2v^\natural_{\rm obj}}+\|\mathbm{O}\oast\mathbm{Y}\|_{\rm F}\right)^2}{\lambda_{\rm min}(\bm G)}\triangleq { \beta},
    \label{eqn:sbound} 
\end{align}
\end{subequations}
in which $v_{\rm obj}^\natural\triangleq v_{\rm obj}(\{\bm{S}_r^\natural,\bm{c}^\natural_r\}_{r=1}^R)$, and $\lambda_{\rm min}(\bm G)>0$ denote the minimal eigenvalue of $\bm G$.
\end{lemma}

\begin{IEEEproof}
    see Appendix \ref{apd:lem_proximal}.
\end{IEEEproof}
The lemma imposes an implicit constraint on any optimal solution of \eqref{eqn:ourformulationExplicit}. Based on the lemma, we define the following:
\begin{Def}[Optimal Solution Set]
The optimal solution set $\mathbb{X}_{\rm sol}$ of \eqref{eqn:ourformulationExplicit} is defined as
\begin{align}
    \mathbb{X}_{\rm sol} &=  \bigg\{ \mathbm{X} ~|~\mathbm{X}  = \sum_{r=1}^R \bm S_r \circ \bm c_r, \nonumber\\
    &\bm{S}_r \geq \bm 0, {\sum_{r=1}^R  \|\bm{S}_r\|_{\rm F}^2}\leq \beta, \bm{c}_r \geq \bm 0, {\sum_{r=1}^R \|\bm{c}_r\|_{\rm F}^2 }\leq \alpha  \bigg\}. 
\end{align}
\end{Def}

To move forward, we will use the following quantity:
\begin{Def}[Sampling-induced Gap]
    The gap induced by $\bm \varOmega$ is defined as
    \begin{align} \label{eq:gap}
        {\rm Gap}^\star(\bm{\bm\varOmega}) = \sup_{{\tilde{\mathbm{X}}} \in \mathbb{X}_{\rm sol}}\left| \frac{\|\mathbm{O}\oast (\mathbm{Y}-{\tilde{\mathbm{X}}})\|_{\rm F}}{\sqrt{|\bm \varOmega|K}} -\frac{ \| \mathbm{Y} - {\tilde{\mathbm{X}}}\|_{\rm F}}{\sqrt{MNK}}\right|.
    \end{align}
\end{Def}
The term $ \rm{Gap}^\star(\bm{\bm\varOmega})$ represents the largest distance between the empirical fitting error over $\bm \varOmega$ and the entire $[M]\times [N]\times [K]$ by any optimal solution of \eqref{eqn:ourformulationExplicit}; see similar definitions in \cite{Wang2012StabilityOM,Shrestha2022DeepSpectrum,Subash2024Quantized}.
Using the definitions and lemmas, we show the following:
\begin{theorem}[Recoverability]\label{thm:recovery}
For any optimal solution of \eqref{eq:ourformulation}, the following inequality holds
\begin{align}
    &\frac{1}{\sqrt{MNK}} \left\| {\mathbm{X}}^\star - \mathbm{X}^\natural \right\|_{\rm F} \nonumber \\
    & \leq \frac{{ \sqrt{v_{\rm obj}^\natural}}}{\sqrt{|\bm \varOmega|K}} + \frac{1}{\sqrt{MNK}} \left\| \mathbm{V}\right\|_{\rm F}+ \rm{Gap}^\star(\bm{\bm\varOmega}),
    \label{eqn:recoverability}
\end{align}
where $\mathbm{V}$ denotes the noise. In addition, with probability $1-\delta$,
\begin{align}
    {\rm Gap}^\star(\bm{\bm\varOmega}) \leq \sqrt{ \frac{\epsilon^2}{|\bm{\varOmega}|} + \frac{\epsilon^2}{MN} + \varepsilon(\bm{\varOmega},\delta,\epsilon)}, \nonumber
\end{align}
where we have
\begin{align}
    {\varepsilon(\bm{\varOmega},\delta,\epsilon)} = {\sqrt{ \Big(\frac{1}{|\bm\varOmega|} - \frac{1}{MN} + \frac{1}{MN|\bm\varOmega|} \Big)  \frac{\xi^2}{2}\log (\frac{2\text{N}(\mathbb{X}_{\rm sol},\epsilon)}{\delta})}},\nonumber
\end{align}
in which $\xi \triangleq \frac{1}{K} (K \iota^2 + \alpha  \beta  )$, $\iota=\| \mathbm{Y}\|_{\infty}$, and $\text{N}(\mathbb{X}_{\rm sol},\epsilon)$ denote the covering number of $\epsilon$-net of $\mathbb{X}_{\rm sol}$.
\end{theorem}
\begin{IEEEproof}
     see Appendix \ref{apd:theorem_recovery}.
\end{IEEEproof}

In the proof of Theorem \ref{thm:recovery}, we used similar ideas from \cite{Shrestha2022DeepSpectrum}. For example, the use of gap term \eqref{eq:gap} to measure the distance between empirical loss and true loss, and the use of covering number to characterize the solution set. 
Theorem \ref{thm:recovery} indicates the optimal solution to \eqref{eq:ourformulation} is close to the ground-truth $\mathbm{X}^\natural$. 
In particular, with a sufficiently large number of samples $|\bm{\varOmega}|$, $\rm{Gap}^\star(\bm{\bm\varOmega})$ in \eqref{eqn:recoverability} can be reasonably small. Additionally, the choice of regularization parameters $\rho$ and $\zeta$ plays an important role in determining the upper bound in \eqref{eqn:recoverability}. With larger $\rho$ and $\zeta$, the solution set of $\bm{S}^\star$ and $\bm{C}^\star$ will be more restricted according to \eqref{eqn:boundedSolution}, leading to a smaller $N(\mathbb{X}_{\rm sol},\epsilon)$ and thus smaller $\rm{Gap}^\star(\bm{\bm\varOmega})$ in \eqref{eqn:recoverability}. However, the first term in \eqref{eqn:recoverability}, which depends on $v_{\rm obj}^\natural$, could increase, as larger $\rho$ and $\zeta$ could lead to overall larger regularization terms in \eqref{eqn:recoverability}.

\subsection{Convergence Analysis}
Convergence of ADMM with PnP denoising has been extensively studied in the literature; see, e.g., \cite{chan2016plug,sreehari2016plug,Teodoro2019convergent,Gavaskar2021OnPnPLinear,ryu2019plug}. However, our formulation uses the PnP denoisers in the latent domain, and the spatio-spectral decomposition makes the problem intrinsically nonconvex.
The nonnegativity constraints on $\bm S_r$ and $\bm c_r$ also make the problem structure more complex than those in existing analyses.
Convergence behaviors under such problem structures require tailored analysis.

In this section, we extend the analysis of \cite{chan2016plug} for data-domain ADMM PnP denoising algorithms and
show that the proposed algorithm achieves fixed point convergence for a broad class of denoisers. 
We should mention that the extension of the proof in \cite{chan2016plug} is not trivial, as handling the PnP denoisers in a latent domain and taking care of the nonnegativity of $\bm S_r$ and $\bm c_r$ brings new complications. As one will see, the special updating rule of HALS plays a key role in establishing convergence under our Algorithm~\ref{alg:admm_pnp}.
In addition, we will show that for linear denoisers satisfying Assumption \ref{assump_linearDenoiser}, the algorithm not only converges to a fixed point, but also reaches a KKT point of Problem \eqref{eq:ourformulation}, a stronger conclusion than that in \cite{chan2016plug}.

Similar as in \cite{chan2016plug}, the following assumptions are adopted:

\begin{assumption} \label{assump_boundDenoiser}
    The denoiser is bounded, i.e., 
\begin{align}
    \| \bm{D}_\sigma(\bm{X}) - \bm{X}\|_{\rm F}^2 /(MN) \leq \sigma^2 C,
\end{align}
where $\bm X\in\mathbb{R}^{M\times N}$,
$\sigma^2 = \lambda/\rho$ denotes the noise power, and $C$ is a constant that is independent of $M$, $N$ and $\sigma$.
\end{assumption}

\begin{assumption} \label{assump_boundGrad}
The partial gradient of the function $f({\{\bm{s}_r, \bm{c}_r\}}_{r=1}^R) = \|  \mathbm{O} \oast \left(\mathbm{Y} - \sum_{r=1}^R\bm{S}_r \circ \bm{c}_r\right) \|_{\rm F}^2$ w.r.t. each $\bm{s}_r={\rm vec}(\bm S_r)$ is bounded; i.e., there exists $L < \infty$, such that 
\begin{align}\label{eqn:boundedDerivative}
    \left\| {\nabla_{\bm s_r} f(\{\bm{s}_r,\bm{c}_r\}_{r=1}^R)} \right\|_2/\sqrt{MN} \leq L.
\end{align}
\end{assumption} 
Since $f(\{\bm{s}_r,\bm{c}_r\}_{r=1}^R)$ is quadratic, its partial gradient w.r.t. $\bm{s}_r$ depends on the empirical error $\mathbm{O} \oast (\mathbm{Y} - \sum_{r=1}^R\bm{S}_r \circ \bm{c}_r)$ and $\bm{c}_r$. Consequently, Assumption \ref{assump_boundGrad} essentially requires that either the empirical error or $\bm c_r$ becomes arbitrarily large in any iteration of Algorithm \ref{alg:admm_pnp}. This is generally a valid assumption, as a large error term will be penalized by $f(\{\bm{s}_r,\bm{c}_r\}_{r=1}^R)$ in \eqref{eq:ourformulation}, while large values of $\bm{c}_r$ will be prevented by $\zeta \bm c_r^\top \bm c_r$.

\medskip
\begin{theorem}[Fixed Point Convergence] \label{theorem_convergence}
Suppose that Assumptions \ref{assump_boundDenoiser} and \ref{assump_boundGrad} hold. Then, the iterates of Algorithm~\ref{alg:admm_pnp} converge to a fixed point. In other words, there exist $\{\barbm{s}_r,\barbm{c}_r,\barbm{z}_r,\barbm{\psi}_r\}_{r=1}^R$ such that $\|\bm{s}_r^{(t)}-\barbm{s}_r\|_2$, $\|\bm{c}_r^{(t)}-\barbm{c}_r\|_2$, $\|\bm{z}_r^{(t)}-\barbm{z}_r\|_2$ and $\|\bm{\psi}_r^{(t)}-\barbm{\psi}_r\|_2$ approach $0$ for all $r\in[R]$ as $t\rightarrow \infty$.
\end{theorem}

\begin{IEEEproof}
    see Appendix \ref{Apd:convergence}.
\end{IEEEproof}

According to Theorem \ref{theorem_convergence}, since $\bm{\psi}_r$ updates as \eqref{eqn:dualupdate} and converges, it implies that $\bm{s}_r^{(t)}$ and $\bm{z}_r^{(t)}$ become identical as $t$ approaches infinity. 
Based on Theorem \ref{theorem_convergence}, stronger guarantees can be provided for linear denoiser cases:

\begin{theorem}[KKT Point Convergence] \label{corollary_KKT}
    Suppose that Assumption \ref{assump_boundGrad} holds, and a linear denoiser satisfying Assumption \ref{assump_linearDenoiser} and \ref{assump_boundDenoiser} is adopted,
    then Algorithm~\ref{alg:admm_pnp} reaches a KKT point of Problem \eqref{eq:ourformulation}.
\end{theorem}

\begin{IEEEproof}
    see Appendix \ref{apd:kkt}.
\end{IEEEproof}

\section{Numerical Results}
\label{sec:exp}

\subsection{Simulation Settings}
\label{subsec:simulationSetting}

\subsubsection{Metrics}
To evaluate the recovery performance, the \textit{relative square error} (RSE) is adopted:
\begin{align}
    \text{RSE}(\widehat{\mathbm{X}}) = {\|\widehat{\mathbm{X}} - {\mathbm{X}^\natural}\|_{\rm F}^2}/\|{{\mathbm{X}^\natural}}\|_{\rm F}^2, \nonumber
\end{align}
in which $\widehat{\mathbm{X}}$ and ${\mathbm{X}^\natural}$ denote the estimated and ground-truth radio maps, respectively.
In addition, we also employ the \textit{structural similarity index measure} (SSIM) \cite{Wang2004SSIM} of the recovered radio maps in the log domain. SSIM reflects similarity perceived by human, and is a score between 0 and 1; the log-transform helps better capture recovery performance in low-power regions. As the radio maps span many frequency bands, the \textit{mean SSIM} (MSSIM) is averaged over $K$ bands.

\subsubsection{Baselines}
We implement two classic interpolation-based methods, i.e., the \textit{thin plate spline} (\texttt{TPS}) \cite{Bazerque2011Splines} and \textit{nearest neighbor interpolation} (\texttt{NN}). 
In addition, we use the \textit{LL1 block-term tensor decomposition based RME method}, namely, \texttt{LL1} \cite{Zhang2020SpectrumViaBlockTerm}, and two deep learning based methods, i.e., \texttt{Nasdac} and \texttt{Dowjons} \cite{Shrestha2022DeepSpectrum}. We also include the data-domain PnP method, which solves \eqref{eqn:RMEproblem_data} using a band-by-band denoising strategy as in \cite{li2024zero}. We refer to this data-domain PnP method as \texttt{DaPnP} (as opposed to our proposed \texttt{LaPnP} method). The results are averaged over $50$ Monte Carlo trials.

\subsubsection{Synthetic Data Generation}
We test our algorithms on two types of simulated radio maps, using a statistical model \cite{goldsmith_2005} and a ray-tracing model (RTM) \cite{yun2015ray,DatasetPaper}, respectively. 
The statistical model (SM) from \cite{goldsmith_2005} captures the probabilistic characteristics of radio propagation under typical scenarios, such as urban or suburban environments. The SM is not for describing instantaneous signal propagation and thus reflects the relatively long-term ``average'' radio environment.
The RTM \cite{yun2015ray} simulates how radio waves propagate by considering instantaneous interactions with objects in the environment. The RTM can incorporate the effects of buildings and other barriers in the region. 
Both models are widely used and considered of interest in the literature; see, e.g., \cite{Zhang2020SpectrumViaBlockTerm,sun2024integrated,Sun2022InterpolationMatrix,Bazerque2011Splines,Subash2024Quantized,Shrestha2022DeepSpectrum,Levie2021radiounet,teganya2021deep,romero2022radio,roger2023deep}.

\noindent
$\bullet$ {\bf Data under SM}: 
In the simulations, we consider a $125 \times 125$m$^2$ region, discretized into $51\times 51$ grids with a grid size of $2.5\times 2.5$m$^2$. The power spectral density of the received signal is measured by the sensors across $32$ discrete frequency bins. Hence, the radio maps have a size of $51\times 51 \times 32$. The ground-truth radio maps are generated as follows:
\begin{align}
    {{\bm{S}}_r^\natural}(m,n) =\frac{10^{\bm{\upsilon}_r(m,n)/10}}{ \| d_0 \times ([m,n]-[m_r,n_r]) \|_2^{ \gamma} },
    \label{eq:SLFgen}
\end{align}
where $d_0=2.5$m, $[m,n]-[m_r,n_r]$ denotes the grid distance between the coordinates $(m,n)$ and that of the $r$th emitter, and $\gamma$ is the path loss exponent randomly selected between $2$ and $2.5$. The term $\bm{\upsilon}_r$ denotes the log-normal shadow fading and follows a Gaussian distribution with zero mean and auto-correlation function $\mathbb{E}[\bm{\upsilon}_r(m_1,n_1)\bm{\upsilon}_r(m_2,n_2)] = \sigma_s^2 \exp(-d_0\|[m_1,n_1]-[m_2,n_2]\|_2/d_\text{c})$, where $\sigma_\text{s}^2$ is the shadowing variance and $d_\text{c}$ denotes the de-correlation distance. A large $\sigma_\text{s}$ and a small $d_\text{c}$ indicate heavy shadowing effects, with typical values of $\sigma_\text{s}$ and $d_\text{c}$ ranging from $4$ to $13$ and $50$ to $100$, respectively.   
The PSDs are generated following the setups in \cite{Shrestha2022DeepSpectrum,Subash2024Quantized}. For each trial, $|\bm \varOmega| = \tau MN$ samples are randomly picked, with $\tau$ denoting the sampling rate.
Under noisy cases, we generate noise $\mathbm{V}$ from zero-mean Gaussian distributions, with the signal-to-noise-ratio (SNR) defined as $\text{SNR} = 10 \log_{10} ({\| \mathbm{X}^\natural \|_{\rm F}^2}/{\|\mathbm{V}\|_{\rm F}^2})$ (dB).

\noindent
$\bullet$ {\bf Data under RTM}: 
Under this simulation setting, the RadioMapSeer dataset \cite{DatasetPaper} is adopted, which includes 701 city maps, each covering a $256 \times 256$m$^2$ area. The maps are further divided into $256 \times 256$ grids.
For simulation efficiency, we use a downsampled version where each map has 128$\times$128 grids.
The dataset considers different numbers of ray-environment interactions, where more interactions provide more details at a cost of increased computational complexity. For our simulation, we adopt the IRT2 setting \cite{DatasetPaper}, which allows each transmit signal (ray) to have up to 2 interactions with the environment (e.g., buildings).
RadioMapSeer is a 2D dataset, where the frequency mode is not considered.
To simulate 3D radio maps, 
we treat the single-emitter maps in RadioMapSeer from the same city as $\bm S_r^\natural$'s,
and generate $\bm c_r^\natural$'s for each emitter. Then we construct the radio maps following \eqref{eqn:X=SC}.

\subsubsection{Algorithm Settings}
For the proposed algorithm, we employ three image denoisers, namely, DSG-NLM \cite{sreehari2016plug}, BM3D \cite{dabov2007image}, and DRUnet \cite{zhang2021plug}. DSG-NLM and BM3D are training-free methods; see Sec. \ref{subsec:pnpDenoise}. DRUnet is a deep denoiser based on the U-Net architecture \cite{ronneberger2015Unet} and is trained on over 8,500 natural images \cite{zhang2021plug}. In particular, for DSG-NLM, we follow the strategy in \cite{sreehari2016plug} and stop updating the denoising matrix $\bm W$ after the 10th iteration of Algorithm \ref{alg:admm_pnp}---which means the denoiser is essentially a linear one after 10 iterations. 
We refer to the proposed method with these denoisers as \texttt{LaPnP-NLM}, \texttt{LaPnP-BM3D}, and \texttt{LaPnP-DRU}, respectively. 
In Algorithm \ref{alg:admm_pnp}, for black-box denoisers (e.g., the deep denoisers), we apply a log-domain transformation to the input data before passing it through the denoisers, and then reverse the transformation on the output. 
These transformations can be considered additional layers of the denoisers.
Such transforms enhance the performance as it ``compresses'' the dynamic range of the input data to the neural denoisers. Note that we do not use such transformations on linear denoisers. 

For the baselines \texttt{Nasdac} and \texttt{Dowjons}, training is performed using 500,000 SLFs generated by \eqref{eq:SLFgen}. Each SLF is generated by uniformly sampling the parameters $\gamma \in [2,2.5]$, $\sigma_\text{s}\in [3,8]$, $d_\text{c} \in [30,100]$ \cite{Shrestha2022DeepSpectrum}. For \texttt{LL1}, we set the number of emitters as the ground truth, and the rank of SLFs as $L=4$, following \cite{Zhang2020SpectrumViaBlockTerm}.

\subsection{Simulation Results Under SM}
\label{subsec:synthetic}

\begin{figure}[t]
    \centering
    \begin{subfigure}{.49\textwidth}
         \centering
         \includegraphics[width=0.95\linewidth]{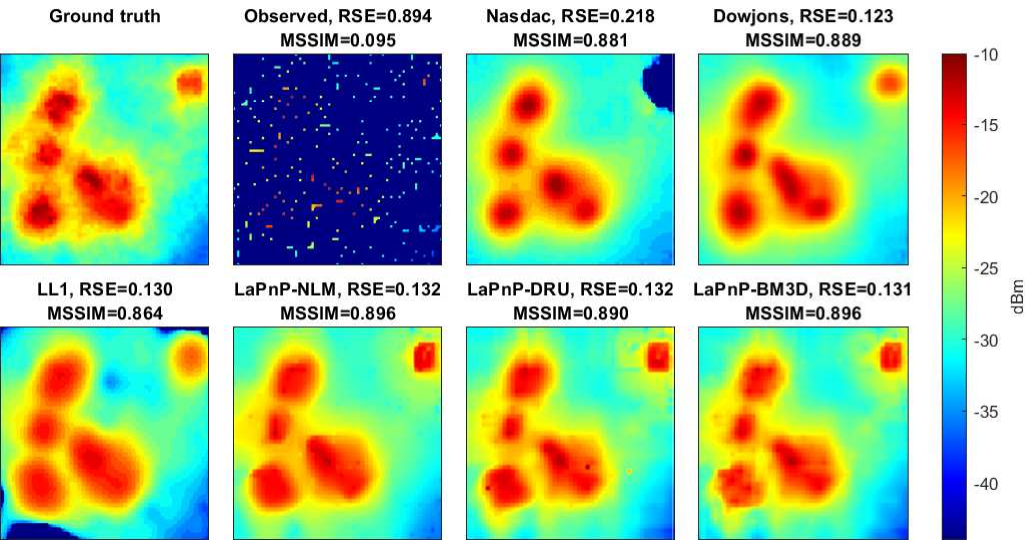}
        \caption{$\sigma_\text{s}=6$, and $R=6$.}
        \label{subfig:visual_moderate}
     \end{subfigure}
     \begin{subfigure}[b]{.49\textwidth}
         \centering
        \includegraphics[width=0.95\linewidth]{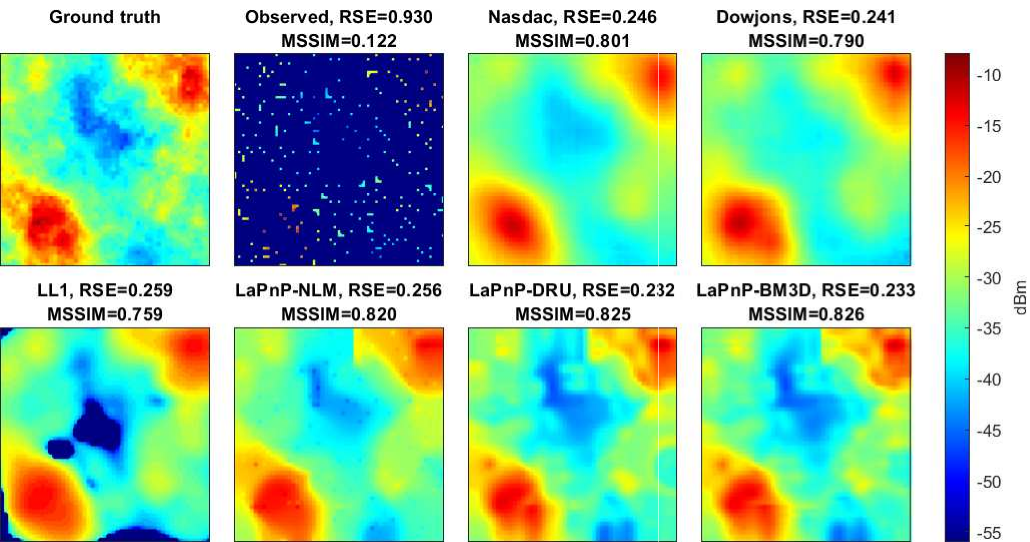}
        \caption{$\sigma_\text{s}=10$, and $R=2$.}
        \label{subfig:visual_highshadowing}
     \end{subfigure}
     \caption{Recovered radio maps under the $10$-th frequency bin; $d_\text{c}=50$, $\tau=10\%$}
     \label{fig:visual}
\end{figure}

Fig. \ref{fig:visual} illustrates the recovered radio maps in the $10$th frequency bin. Fig. \ref{subfig:visual_moderate} shows results with $\sigma_\text{s}=6$ and $R=6$. All methods produce visually reasonable recoveries, except for \texttt{Nasdac}, which fails to capture the high energy pattern in the top-right corner. Note that the \texttt{LaPnP} family does not need training on SLFs (as opposed to \texttt{DowJons} and \texttt{Nasdac}), yet the performance is rather competitive.
Fig. \ref{subfig:visual_highshadowing} illustrates a more challenging heavy shadowing scenario, with $\sigma_\text{s}=10$ and $R=2$. As one can see, all three \texttt{LaPnP} variants recover accurate contours for both high power and low power regions. In contrast, other methods tend to produce overly smoothed radio maps and fail to recover the low power region effectively. 
For \texttt{Nasdac} and \texttt{Dowjons}, note that the training data are generated with $\sigma_\text{s} \leq 8$ \cite{Shrestha2022DeepSpectrum}. As $\sigma_\text{s}$ exceeds $8$ in this simulation, they struggle to capture detailed characteristics of the radio map. 
As mentioned, such training-testing mismatches can cause performance deterioration of these learning-based methods, but our PnP based method does not have this issue.

\begin{table}[!tb]
\caption{Performance under different $\tau$; $\sigma_\text{s}=6$, $d_c=50$, $R = 6$.} 
\footnotesize
\centering
\scalebox{0.76}{
\begin{tabular}{| c | c c c c c c| }
\hline
 & \multicolumn{6}{c|}{RSE}\\
$\tau$ &  \texttt{Nasdac} & \texttt{Dowjons}  & \texttt{LL1} & \texttt{LaPnP-NLM}  & \texttt{LaPnP-DRU} & \texttt{LaPnP-BM3D} \\ 
 \hline
5\% & 0.283 & \textbf{0.184} & 0.243 & 0.279 & 0.240 & 0.219 \\
10\% & 0.221 & 0.130 & 0.138 & 0.151 & 0.137 & \textbf{0.129} \\
15\% & 0.209 & 0.097 & 0.103 & 0.104 & 0.092 & \textbf{0.089} \\
20\% & 0.201 & 0.080 & 0.079 & 0.078 & 0.073 & \textbf{0.069} \\
\hline
\hline
 & \multicolumn{6}{c|}{MSSIM}\\
$\tau$ &  \texttt{Nasdac} & \texttt{Dowjons}  & \texttt{LL1} & \texttt{LaPnP-NLM}  & \texttt{LaPnP-DRU} & \texttt{LaPnP-BM3D} \\ 
 \hline
5\% & 0.8085 & 0.8391 & 0.7617 & 0.8233 & 0.8166 & \textbf{0.8442} \\
10\% & 0.8248 & 0.8504 & 0.8433 & 0.8725 & 0.8640 & \textbf{0.8869} \\
15\% & 0.8262 & 0.8529 & 0.8553 & 0.8922 & 0.8848 & \textbf{0.9033} \\
20\% & 0.8223 & 0.8544 & 0.8703 & 0.9046 & 0.9022 & \textbf{0.9165} \\
 \hline
\end{tabular}}
\label{table:perf_mr}
\end{table}

\begin{table}[!tb]
\caption{Performance under different $\sigma_\text{s}$; {$d_c=50$, $\tau=10\%$ and $R=6$.}}
\footnotesize
\centering
\scalebox{0.76}{
\begin{tabular}{ | c | c c c c c c | }
\hline
 & \multicolumn{6}{c|}{RSE}\\
$\sigma_{\text{s}}$ &  \texttt{Nasdac} & \texttt{Dowjons}  & \texttt{LL1} & \texttt{LaPnP-NLM}  & \texttt{LaPnP-DRU} & \texttt{LaPnP-BM3D} \\ 
\hline
4 & 0.194 & \textbf{0.086} & 0.110 & 0.129 & 0.091 & 0.092 \\
6 & 0.241 & \textbf{0.125} & 0.140 & 0.159 & 0.130 & 0.125 \\
8 & 0.267 & 0.165 & 0.178 & 0.194 & 0.174 & \textbf{0.162} \\
10 & 0.309 & 0.222 & 0.224 & 0.235 & 0.223 & \textbf{0.209} \\
12 & 0.338 & 0.292 & 0.270 & 0.275 & 0.284 & \textbf{0.259} \\
14 & 0.387 & 0.356 & 0.328 & 0.329 & 0.335 & \textbf{0.309} \\
\hline
\hline
 & \multicolumn{6}{c|}{MSSIM}\\
$\sigma_{\text{s}}$ &  \texttt{Nasdac} & \texttt{Dowjons}  & \texttt{LL1} & \texttt{LaPnP-NLM}  & \texttt{LaPnP-DRU} & \texttt{LaPnP-BM3D} \\ 
\hline
4 & 0.8693 & 0.9006 & 0.8964 & 0.9157 & 0.9179 & \textbf{0.9307} \\
6 & 0.8136 & 0.8509 & 0.8549 & 0.8726 & 0.8774 & \textbf{0.8869} \\
8 & 0.7633 & 0.7956 & 0.7965 & 0.8226 & 0.8254 & \textbf{0.8413} \\
10 & 0.6943 & 0.7277 & 0.7319 & 0.7654 & 0.7733 & \textbf{0.7893} \\
12 & 0.6570 & 0.6884 & 0.6727 & 0.7388 & 0.7403 & \textbf{0.7607} \\
14 & 0.5959 & 0.6264 & 0.5969 & 0.6947 & 0.6996 & \textbf{0.7210} \\
 \hline
\end{tabular}}
\label{table:perf_shadow}
\end{table}

\begin{table}[!tb]
\caption{Performance under different noise; $\sigma_\text{s}=6$, $d_\text{c}=50$, $\tau = 10\%$, and $R = 6$.} 
\footnotesize
\centering
\scalebox{0.76}{
\begin{tabular}{| c | c c c c c c |}
\hline
 & \multicolumn{6}{c|}{RSE}\\
SNR/dB & \shortstack{\texttt{DaPnP-} \\ \texttt{NLM}} &\shortstack{\texttt{DaPnP-} \\ \texttt{DRU}}  & \shortstack{\texttt{DaPnP-} \\ \texttt{BM3D}} & \shortstack{\texttt{LaPnP-} \\ \texttt{NLM}} &\shortstack{\texttt{LaPnP-} \\ \texttt{DRU}}  & \shortstack{\texttt{LaPnP-} \\ \texttt{BM3D}}\\ 
\hline
clean & 0.154 & 0.145 & 0.152 & 0.151 & 0.137 & \textbf{0.129}\\
20& 0.162 & 0.147 & 0.156 & 0.158 & 0.140 & \textbf{0.131}\\
10 & 0.173 & 0.159 & 0.163 & 0.165 & 0.152 & \textbf{0.137}\\
\hline
\hline
 & \multicolumn{6}{c|}{MSSIM}\\
SNR/dB & \shortstack{\texttt{DaPnP-} \\ \texttt{NLM}} &\shortstack{\texttt{DaPnP-} \\ \texttt{DRU}}  & \shortstack{\texttt{DaPnP-} \\ \texttt{BM3D}} & \shortstack{\texttt{LaPnP-} \\ \texttt{NLM}} &\shortstack{\texttt{LaPnP-} \\ \texttt{DRU}}  & \shortstack{\texttt{LaPnP-} \\ \texttt{BM3D}}\\ 
\hline
clean & 0.8794 & \textbf{0.8902} & 0.8858 & 0.8725 & 0.8640 & 0.8869 \\
20&  0.4979 & 0.5702 & 0.5179 & 0.8301 & 0.8286 & \textbf{0.8382}\\
10 & 0.2118 & 0.3013 & 0.2274 & 0.7719 & 0.7603 & \textbf{0.7853}\\
 \hline
 \hline
Runtime/s &112.96 & 251.93 & 8.86 & 21.69 & 52.90 & 2.21 \\
\hline
\end{tabular}}
\label{table:perf_SNR}
\end{table}

Table \ref{table:perf_mr} shows the performance of all methods under various sampling rates $\tau$. One can see that \texttt{LaPnP-BM3D} achieves the overall best performance. Moreover, \texttt{LaPnP-NLM} and \texttt{LaPnP-DRU} perform similarly to the best non-PnP method, namely, \texttt{Dowjons}. However, it is worth noting that \texttt{Dowjons} is a deep learning based method trained on 500,000 SLFs, while the proposed method does not need any training.

Table \ref{table:perf_shadow} summarizes the results under different shadowing variance $\sigma_\text{s}^2$. All three \texttt{LaPnP} variants show competitive RSEs and consistently higher MSSIM values compared to other methods. Notably, \texttt{LaPnP-BM3D} achieves the best performance in terms of both RSE and MSSIM for $\sigma_\text{s}\geq 6$, which indicates moderate to heavy shadowing effects. In particular, as $\sigma_\text{s}$ increases, the MSSIM gap between \texttt{LaPnP}-based methods and the others becomes larger. Again, this happens for the deep learning-based methods because of the model mismatches (i.e., \texttt{Nasdac} and \texttt{Dowjons} used $\sigma_\text{s}\leq8$ during training). The \texttt{LL1} method does not work well for large $\sigma_s$ as the low-rank assumption on $\bm S_r$ no longer holds when shadowing increases.

Table \ref{table:perf_SNR} compares the performance of \texttt{DaPnP} and \texttt{LaPnP} under different SNRs.  When there is no noise, \texttt{DaPnP}-based methods exhibit comparable performance to their \texttt{LaPnP} counterparts. 
Under noisy cases, while all methods show performance degradation, \texttt{DaPnP}-based methods are significantly more affected relative to others. \texttt{LaPnP} deals with all bands simultaneously by using the decomposition model \eqref{eqn:X=SC}, which is similar to PCA and thus is inherently more robust to noise.
In addition, \texttt{DaPnP}'s computational complexity is significantly higher---all three \texttt{DaPnP} variants take 4-5.2 times longer than the corresponding \texttt{LaPnP} versions.

\begin{figure}[t]
    \centering
      \centering
      \includegraphics[width=0.95\linewidth]{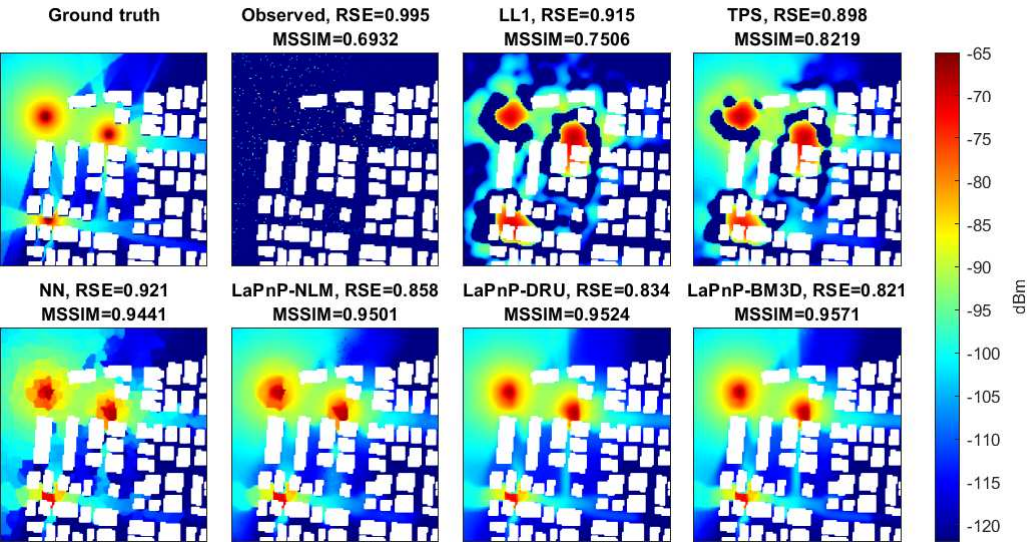}
    \caption{Recovered radio maps in the $10$-th frequency bin; $\tau=5\%$, $R=3$.}
    \label{fig:visual_seer}
\end{figure}

\begin{table}[!tb]
\caption{Performance under RTM and different $\tau$; $R=3$.} 
\footnotesize
\centering
\scalebox{0.76}{
\begin{tabular}{ |c | c c c c c c |}
\hline
 & \multicolumn{6}{c|}{RSE}\\
$\tau$ & \texttt{LL1} & \texttt{TPS}  & \texttt{NN} & \texttt{LaPnP-NLM}  & \texttt{LaPnP-DRU} & \texttt{LaPnP-BM3D} \\ 
 \hline
5\% & 0.911 & 0.907 & 0.979 & 0.907 & 0.873 & \textbf{0.871} \\
10\% & 0.782 & 0.777 & 0.807 & 0.689 & 0.680 & \textbf{0.634} \\
15\% & 0.725 & 0.726 & 0.771 & 0.625 & 0.622 & \textbf{0.582} \\
20\% & 0.620 & 0.612 & 0.606 & 0.564 & \textbf{0.428} & 0.452 \\
\hline
\hline
 & \multicolumn{6}{c|}{MSSIM}\\
$\tau$ &  \texttt{LL1} & \texttt{TPS}  & \texttt{NN} & \texttt{LaPnP-NLM}  & \texttt{LaPnP-DRU} & \texttt{LaPnP-BM3D} \\ 
 \hline
5\% & 0.8224 & 0.8692 & 0.9395 & 0.9399 & 0.9426 & \textbf{0.9464} \\
10\% & 0.8277 & 0.8824 & 0.9555 & 0.9566 & \textbf{0.9615} & 0.9572 \\
15\% & 0.8335 & 0.8992 & 0.9649 & 0.9648 & \textbf{0.9699} & 0.9682 \\
20\% & 0.8406 & 0.9110 & 0.9695 & 0.9688 & 0.9737 & \textbf{0.9741} \\
 \hline
\end{tabular}}
\label{table:perf_mr_seer}
\end{table}

\begin{figure}[t!]
    \centering
    \begin{subfigure}{0.47\linewidth}
      \centering
      \includegraphics[width=\linewidth]{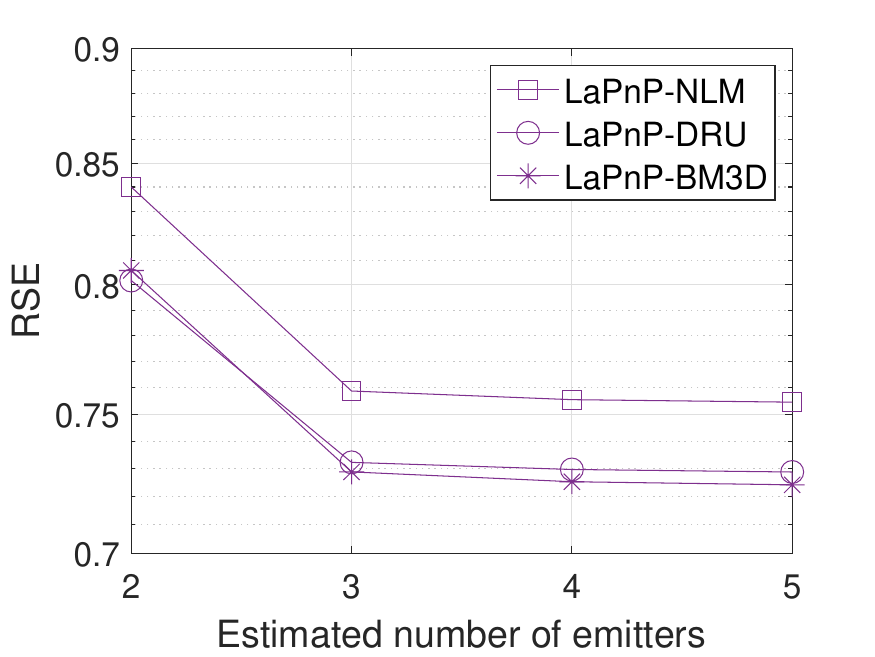}
      \caption{RSE}
      \label{subfig:rse_rest}
    \end{subfigure}
    \begin{subfigure}{0.47\linewidth}
      \centering
      \includegraphics[width=\linewidth]{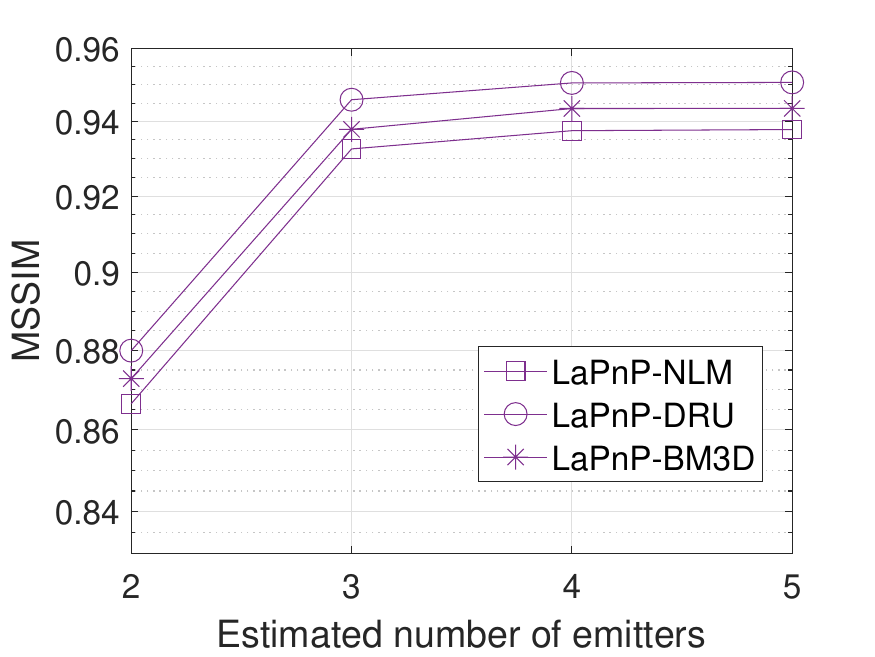}  
      \caption{MSSIM}
      \label{subfig:ssim_rest}
    \end{subfigure}
    \caption{Performance under RTM and different estimated $\hat{R}$'s; $\tau=10\%$, $R = 3$.}
    \label{fig:perf_Rest}
\end{figure}

\subsection{Simulation results under RTM}
Fig. \ref{fig:visual_seer} shows the result under a RadioMapSeer map at the $10$th band. All \texttt{LaPnP} variants achieve more accurate reconstruction. The baselines \texttt{LL1} and \texttt{TPS} mistakenly introduce non-existing signal power in low-energy regions, while \texttt{NN} inaccurately reconstructs the high-energy regions with noticeable blocking artifacts.
Note that \texttt{DowJons} and \texttt{Nasdac}'s training mechanism requires nontrivial modifications to incorporate the building maps. 
Hence, these two baselines are not included in the RTM simulations.

Table \ref{table:perf_mr_seer} presents the performance under various sampling rates. 
The results are averaged over 50 randomly selected city maps.
One can see that \texttt{LaPnP-BM3D} and \texttt{LaPnP-DRUnet} consistently deliver superior results. Notably, they obtain significantly higher MSSIMs compared to those of \texttt{LL1} and \texttt{TPS}.

Fig. \ref{fig:perf_Rest} evaluates the impact of incorrectly estimating $R$ onto the proposed method, where the ground-truth $R$ is $3$. As one can see, underestimating $R$ leads to a noticeable drop in performance for all \texttt{LaPnP} variants, which is understandable as it leads to information loss in the decomposition model. However, when $R$ is overestimated, the performance remains relatively stable.

\subsection{Real-World Data Experiment}
\label{sec:realEXP}
This real-world dataset \cite{compass2008} collects the signal strength across 9 frequency bands in a $14\times34$m$^2$ indoor area of Mannheim University. The area is divided into $1\times1$m$^2$ grid cells, resulting in the radio map of size $14\times 34\times 9$. In 166 of the grid cells, sensors are deployed, measuring the PSDs in all 9 frequency bins.
More details of this dataset can be found in \cite{Zhang2020SpectrumViaBlockTerm,Shrestha2022DeepSpectrum,compass2008}.

\begin{figure}[!tb]
    \centering
    \includegraphics[width=0.95\linewidth]{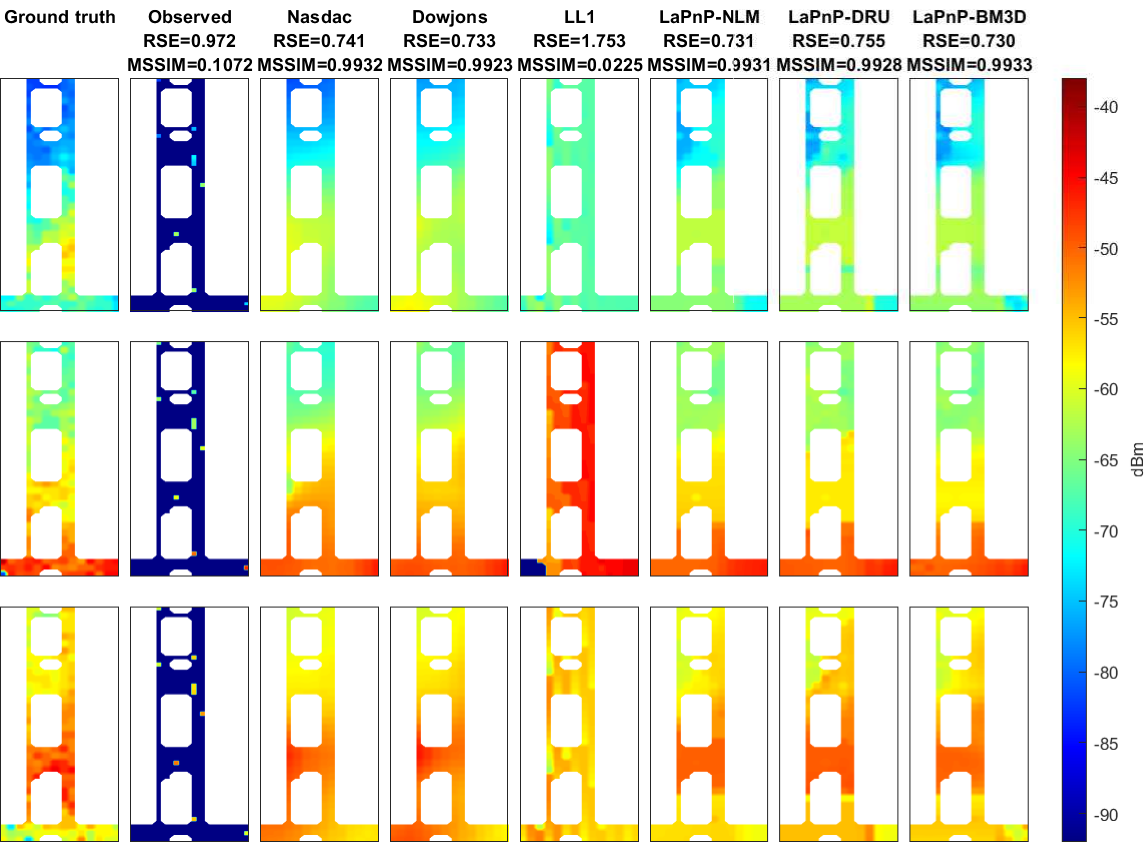}
    \caption{Recovered radio maps in the $1$st, $5$th and $9$th band from the Mannheim data experiment; $\tau$ = 5\%.}
    \label{fig:mannheim}
\end{figure}

Fig. \ref{fig:mannheim} shows the 1st, 5th and 9th bands of the recovered Mannheim radio map under $\tau=5\%$. The number of emitters is set as $R=7$ for all methods, following \cite{Shrestha2022DeepSpectrum,Subash2024Quantized}.
The baselines \texttt{Nasdac} and \texttt{Dowjons} are trained as before, and the setup of \texttt{LL1} follows that in \cite{Zhang2020SpectrumViaBlockTerm}.
As one can see, deep learning-based methods and the proposed methods output visually similar results. The \texttt{LL1} completely fails under such a low sampling rate. Table \ref{table:perf_mr_mannheim} shows the performance metrics under different sampling rates. The proposed methods outperform others in terms of RSEs, and achieve similar MSSIMs as those of \texttt{Dowjons}. Again, we stress that the proposed method does not need any radio map training data that matches the environment of the Mannheim data.

\begin{table}[!tb]
\caption{Performance under different sampling rate, Mannheim dataset.} 
\footnotesize
\centering
\scalebox{0.76}{
\begin{tabular}{| c | c c c c c c |}
\hline
 & \multicolumn{6}{c|}{RSE}\\
$\tau$ &  \texttt{Nasdac} & \texttt{Dowjons}  & \texttt{LL1} & \texttt{LaPnP-NLM}  & \texttt{LaPnP-DRU} & \texttt{LaPnP-BM3D} \\ 
 \hline
5\%& 0.812 & \textbf{0.756} & 2.065 & 0.783 & 0.807 & 0.764 \\
10\%& 0.595 & 0.572 & 2.033 & 0.642 & 0.681 & \textbf{0.567} \\
15\%& 0.564 & 0.540 & 1.935 & 0.535 & 0.581 & \textbf{0.499} \\
20\%& 0.504 & 0.423 & 2.097 & 0.445 & 0.469 & \textbf{0.412} \\
\hline
\hline
 & \multicolumn{6}{c|}{MSSIM}\\
$\tau$ &  \texttt{Nasdac} & \texttt{Dowjons}  & \texttt{LL1} & \texttt{LaPnP-NLM}  & \texttt{LaPnP-DRU} & \texttt{LaPnP-BM3D} \\ 
 \hline
5\%& 0.9880 &  0.9922 & 0.7346 & 0.9932 & 0.9928 & \textbf{0.9933} \\
10\%& 0.9933 & 0.9927 & 0.7478 & 0.9931 & 0.9919 & \textbf{0.9936} \\
15\%& 0.9945 & 0.9939 & 0.7543 & \textbf{0.9951} & 0.9934 & 0.9943 \\
20\%& \textbf{0.9964} & 0.9947 & 0.7481 & 0.9957 & 0.9950 & 0.9953 \\
 \hline
\end{tabular}}
\label{table:perf_mr_mannheim}
\end{table}

\section{Conclusion}
This paper revisited the RME problem and proposed an ADMM PnP denoising based algorithm. The algorithm leverages well-developed natural image denoisers to impose implicit structural constraints on radio maps. This way, expensive training using large amounts of radio maps can be circumvented, yet complex and intricate structural information of signal denoising processes captured by sophisticated denoising functions is still exploited.
The method also does not suffer from training-testing mismatches as traditional deep learning based RME methods do.
Unlike conventional PnP denoising methods that are usually applied in the data domain, our method ``embeds'' the denoisers in the latent spatial domain, reducing computational complexity and improving noise robustness.
Critical theoretical aspects such as the recoverability of the radio maps and convergence of the ADMM algorithm were also studied, providing performance characterizations under reasonable conditions. Simulations and real-world data experiments were presented to validate our algorithm.

\bigskip

The code of our algorithm is available at \url{https://github.com/xumaomao94/LaPnP}. The Mannheim University dataset is available at \url{https://ieee-dataport.org/open-access/crawdad-mannheimcompass-v-2008-04-11}.

\bibliographystyle{IEEEtran}

\begin{thebibliography}{10}
\providecommand{\url}[1]{#1}
\csname url@samestyle\endcsname
\providecommand{\newblock}{\relax}
\providecommand{\bibinfo}[2]{#2}
\providecommand{\BIBentrySTDinterwordspacing}{\spaceskip=0pt\relax}
\providecommand{\BIBentryALTinterwordstretchfactor}{4}
\providecommand{\BIBentryALTinterwordspacing}{\spaceskip=\fontdimen2\font plus
\BIBentryALTinterwordstretchfactor\fontdimen3\font minus \fontdimen4\font\relax}
\providecommand{\BIBforeignlanguage}[2]{{%
\expandafter\ifx\csname l@#1\endcsname\relax
\typeout{** WARNING: IEEEtran.bst: No hyphenation pattern has been}%
\typeout{** loaded for the language `#1'. Using the pattern for}%
\typeout{** the default language instead.}%
\else
\language=\csname l@#1\endcsname
\fi
#2}}
\providecommand{\BIBdecl}{\relax}
\BIBdecl

\bibitem{zeng2024tutorial}
Y.~Zeng, J.~Chen, J.~Xu, D.~Wu, X.~Xu, S.~Jin, X.~Gao, D.~Gesbert, S.~Cui, and R.~Zhang, ``A tutorial on environment-aware communications via channel knowledge map for 6{G},'' \emph{IEEE Commun. Surveys Tuts.}, 2024.

\bibitem{Bi2019radiomap}
S.~Bi, J.~Lyu, Z.~Ding, and R.~Zhang, ``Engineering radio maps for wireless resource management,'' \emph{IEEE Wireless Commun.}, vol.~26, no.~2, pp. 133--141, 2019.

\bibitem{romero2022radio}
D.~Romero and S.-J. Kim, ``Radio map estimation: A data-driven approach to spectrum cartography,'' \emph{IEEE Signal Process. Mag.}, vol.~39, no.~6, pp. 53--72, 2022.

\bibitem{Bazerque2011Splines}
J.~A. Bazerque, G.~Mateos, and G.~B. Giannakis, ``Group-lasso on splines for spectrum cartography,'' \emph{IEEE Trans. Signal Process.}, vol.~59, no.~10, pp. 4648--4663, 2011.

\bibitem{kim2010cooperative}
S.-J. Kim, E.~Dall'Anese, and G.~B. Giannakis, ``Cooperative spectrum sensing for cognitive radios using kriged kalman filtering,'' \emph{IEEE J. Sel. Topics Signal Process.}, vol.~5, no.~1, pp. 24--36, 2010.

\bibitem{Boccolini2012Kriging}
G.~Boccolini, G.~Hernández-Peñaloza, and B.~Beferull-Lozano, ``Wireless sensor network for spectrum cartography based on kriging interpolation,'' in \emph{Proc.\ IEEE International Symposium on Personal, Indoor and Mobile Radio Communications (PIMRC)}, 2012.

\bibitem{Zhang2020SpectrumViaBlockTerm}
G.~Zhang, X.~Fu, J.~Wang, X.-L. Zhao, and M.~Hong, ``Spectrum cartography via coupled block-term tensor decomposition,'' \emph{IEEE Trans. Signal Process.}, vol.~68, pp. 3660--3675, 2020.

\bibitem{sun2024integrated}
H.~Sun and J.~Chen, ``Integrated interpolation and block-term tensor decomposition for spectrum map construction,'' \emph{IEEE Trans. Signal Process.}, vol.~72, pp. 3896--3911, 2024.

\bibitem{Sun2022InterpolationMatrix}
------, ``Propagation map reconstruction via interpolation assisted matrix completion,'' \emph{IEEE Trans. Signal Process.}, vol.~70, pp. 6154--6169, 2022.

\bibitem{Shrestha2022DeepSpectrum}
S.~Shrestha, X.~Fu, and M.~Hong, ``Deep spectrum cartography: completing radio map tensors using learned neural models,'' \emph{IEEE Trans. Signal Process.}, vol.~70, pp. 1170--1184, 2022.

\bibitem{Subash2024Quantized}
S.~Timilsina, S.~Shrestha, and X.~Fu, ``Quantized radio map estimation using tensor and deep generative models,'' \emph{IEEE Trans. Signal Process.}, vol.~72, pp. 173--189, 2024.

\bibitem{Levie2021radiounet}
R.~Levie, {\c C}.~Yapar, G.~Kutyniok, and G.~Caire, ``Radio{U}net: Fast radio map estimation with convolutional neural networks,'' \emph{IEEE Trans. Wireless Commun.}, vol.~20, no.~6, pp. 4001--4015, 2021.

\bibitem{teganya2021deep}
Y.~Teganya and D.~Romero, ``Deep completion autoencoders for radio map estimation,'' \emph{IEEE Trans. Wireless Commun.}, vol.~21, no.~3, pp. 1710--1724, 2021.

\bibitem{roger2023deep}
S.~Roger, M.~Brambilla, B.~C. Tedeschini, C.~Botella-Mascarell, M.~Cobos, and M.~Nicoli, ``Deep-learning-based radio map reconstruction for {V}2{X} communications,'' \emph{IEEE Trans. Veh. Technol.}, vol.~73, no.~3, pp. 3863--3871, 2024.

\bibitem{Kamilov2023PNP}
U.~S. Kamilov, C.~A. Bouman, G.~T. Buzzard, and B.~Wohlberg, ``Plug-and-play methods for integrating physical and learned models in computational imaging: Theory, algorithms, and applications,'' \emph{IEEE Signal Process. Mag.}, vol.~40, no.~1, pp. 85--97, 2023.

\bibitem{venkatakrishnan2013plug}
S.~V. Venkatakrishnan, C.~A. Bouman, and B.~Wohlberg, ``Plug-and-play priors for model based reconstruction,'' in \emph{Proc.\ IEEE Global Conference on Signal and Information Processing (GlobalSIP)}, 2013.

\bibitem{li2024zero}
S.~Li, L.~Cheng, X.~Fu, and J.~Li, ``Zero-shot reconstruction of ocean sound speed field tensors: A deep plug-and-play approach,'' \emph{The Journal of the Acoustical Society of America}, vol. 155, no.~5, pp. 3475--3489, 2024.

\bibitem{Liu2022HSIpnp}
Y.-Y. Liu, X.-L. Zhao, Y.-B. Zheng, T.-H. Ma, and H.~Zhang, ``Hyperspectral image restoration by tensor fibered rank constrained optimization and plug-and-play regularization,'' \emph{IEEE Trans. Geosci. Remote Sens.}, vol.~60, pp. 1--17, 2022.

\bibitem{wei2020tuning}
K.~Wei, A.~Aviles-Rivero, J.~Liang, Y.~Fu, C.-B. Sch{\"o}nlieb, and H.~Huang, ``Tuning-free plug-and-play proximal algorithm for inverse imaging problems,'' in \emph{Proc. International Conference on Machine Learning (ICML)}.\hskip 1em plus 0.5em minus 0.4em\relax PMLR, 2020.

\bibitem{liu2020rare}
J.~Liu, Y.~Sun, C.~Eldeniz, W.~Gan, H.~An, and U.~S. Kamilov, ``Rare: Image reconstruction using deep priors learned without groundtruth,'' \emph{IEEE J. Sel. Topics Signal Process.}, vol.~14, no.~6, pp. 1088--1099, 2020.

\bibitem{buades2005non}
A.~Buades, B.~Coll, and J.-M. Morel, ``A non-local algorithm for image denoising,'' in \emph{Proc.\ IEEE Conference on Computer Vision and Pattern Recognition (CVPR)}.\hskip 1em plus 0.5em minus 0.4em\relax IEEE, 2005.

\bibitem{dabov2007image}
K.~Dabov, A.~Foi, V.~Katkovnik, and K.~Egiazarian, ``Image denoising by sparse 3-{D} transform-domain collaborative filtering,'' \emph{IEEE Trans. Image Process.}, vol.~16, no.~8, pp. 2080--2095, 2007.

\bibitem{zhang2018ffdnet}
K.~Zhang, W.~Zuo, and L.~Zhang, ``{FFDN}et: Toward a fast and flexible solution for {C}{N}{N}-based image denoising,'' \emph{IEEE Trans. Image Process.}, vol.~27, no.~9, pp. 4608--4622, 2018.

\bibitem{zhang2021plug}
K.~Zhang, Y.~Li, W.~Zuo, L.~Zhang, L.~Van~Gool, and R.~Timofte, ``Plug-and-play image restoration with deep denoiser prior,'' \emph{IEEE Trans. Pattern Anal. Mach. Intell.}, vol.~44, no.~10, pp. 6360--6376, 2021.

\bibitem{chan2016plug}
S.~H. Chan, X.~Wang, and O.~A. Elgendy, ``Plug-and-play {ADMM} for image restoration: Fixed-point convergence and applications,'' \emph{IEEE Trans. Comput. Imag.}, vol.~3, no.~1, pp. 84--98, 2016.

\bibitem{ryu2019plug}
E.~Ryu, J.~Liu, S.~Wang, X.~Chen, Z.~Wang, and W.~Yin, ``Plug-and-play methods provably converge with properly trained denoisers,'' in \emph{Proc. International Conference on Machine Learning (ICML)}.\hskip 1em plus 0.5em minus 0.4em\relax PMLR, 2019.

\bibitem{sreehari2016plug}
S.~Sreehari, S.~V. Venkatakrishnan, B.~Wohlberg, G.~T. Buzzard, L.~F. Drummy, J.~P. Simmons, and C.~A. Bouman, ``Plug-and-play priors for bright field electron tomography and sparse interpolation,'' \emph{IEEE Trans. Comput. Imag.}, vol.~2, no.~4, pp. 408--423, 2016.

\bibitem{Teodoro2019convergent}
A.~M. Teodoro, J.~M. Bioucas-Dias, and M.~A.~T. Figueiredo, ``A convergent image fusion algorithm using scene-adapted {G}aussian-mixture-based denoising,'' \emph{IEEE Trans. Image Process.}, vol.~28, no.~1, pp. 451--463, 2019.

\bibitem{bazerque2009distributed}
J.~A. Bazerque and G.~B. Giannakis, ``Distributed spectrum sensing for cognitive radio networks by exploiting sparsity,'' \emph{IEEE Trans. Signal Process.}, vol.~58, no.~3, pp. 1847--1862, 2009.

\bibitem{romero2017learning}
D.~Romero, S.-J. Kim, G.~B. Giannakis, and R.~L{\'o}pez-Valcarce, ``Learning power spectrum maps from quantized power measurements,'' \emph{IEEE Trans. Signal Process.}, vol.~65, no.~10, pp. 2547--2560, 2017.

\bibitem{Lee2017CGMcartography}
D.~Lee, S.-J. Kim, and G.~B. Giannakis, ``Channel gain cartography for cognitive radios leveraging low rank and sparsity,'' \emph{IEEE Trans. Wireless Commun.}, vol.~16, no.~9, pp. 5953--5966, 2017.

\bibitem{GAVASKAR2023109100}
R.~G. Gavaskar, C.~D. Athalye, and K.~N. Chaudhury, ``On exact and robust recovery for plug-and-play compressed sensing,'' \emph{Signal Processing}, vol. 211, p. 109100, 2023.

\bibitem{Gavaskar2021OnPnPLinear}
------, ``On plug-and-play regularization using linear denoisers,'' \emph{IEEE Trans. Image Process.}, vol.~30, pp. 4802--4813, 2021.

\bibitem{chan2019PnPGraph}
S.~H. Chan, ``Performance analysis of plug-and-play {ADMM}: A graph signal processing perspective,'' \emph{IEEE Trans. Comput. Imag.}, vol.~5, no.~2, pp. 274--286, 2019.

\bibitem{liu2021recovery}
J.~Liu, S.~Asif, B.~Wohlberg, and U.~Kamilov, ``Recovery analysis for plug-and-play priors using the restricted eigenvalue condition,'' in \emph{Proc.\ Advances in Neural Information Processing Systems (NeuIPS)}, 2021.

\bibitem{Xu2025RME}
L.~Xu, L.~Cheng, J.~Chen, W.~Pu, and X.~Fu, ``Radio map estimation via latent-domain plug-and-play denoisers,'' in \emph{Proc.\ IEEE International Conference on Acoustics, Speech and Signal Processing (ICASSP)}, 2025.

\bibitem{goldsmith_2005}
A.~Goldsmith, \emph{Wireless Communications}.\hskip 1em plus 0.5em minus 0.4em\relax Cambridge University Press, 2005.

\bibitem{sionna2022}
J.~Hoydis, S.~Cammerer, F.~{Ait Aoudia}, A.~Vem, N.~Binder, G.~Marcus, and A.~Keller, ``Sionna: An open-source library for next-generation physical layer research,'' \emph{arXiv preprint}, Mar. 2022.

\bibitem{chouvardas2016method}
S.~Chouvardas, S.~Valentin, M.~Draief, and M.~Leconte, ``A method to reconstruct coverage loss maps based on matrix completion and adaptive sampling,'' in \emph{Proc.\ IEEE International Conference on Acoustics, Speech and Signal Processing (ICASSP)}, 2016.

\bibitem{Boyd2011ADMM}
S.~Boyd, N.~Parikh, E.~Chu, B.~Peleato, J.~Eckstein \emph{et~al.}, ``Distributed optimization and statistical learning via the alternating direction method of multipliers,'' \emph{Foundations and Trends{\textregistered} in Machine Learning}, vol.~3, no.~1, pp. 1--122, 2011.

\bibitem{Majee2021CT}
S.~Majee, T.~Balke, C.~A.~J. Kemp, G.~T. Buzzard, and C.~A. Bouman, ``Multi-slice fusion for sparse-view and limited-angle 4d ct reconstruction,'' \emph{IEEE Tran. Comput. Imag.}, vol.~7, pp. 448--462, 2021.

\bibitem{Milanfar2013tour}
P.~Milanfar, ``A tour of modern image filtering: New insights and methods, both practical and theoretical,'' \emph{IEEE Signal Process. Mag.}, vol.~30, no.~1, pp. 106--128, 2013.

\bibitem{szeliski2022computer}
R.~Szeliski, \emph{Computer vision: algorithms and applications}.\hskip 1em plus 0.5em minus 0.4em\relax Springer Nature, 2022.

\bibitem{yu2011solving}
G.~Yu, G.~Sapiro, and S.~Mallat, ``Solving inverse problems with piecewise linear estimators: From gaussian mixture models to structured sparsity,'' \emph{IEEE Trans. Image Process.}, vol.~21, no.~5, pp. 2481--2499, 2011.

\bibitem{milanfar2024denoising}
P.~Milanfar and M.~Delbracio, ``Denoising: A powerful building-block for imaging, inverse problems, and machine learning,'' \emph{arXiv preprint arXiv:2409.06219}, 2024.

\bibitem{zhou2014nonnegative}
G.~Zhou, A.~Cichocki, Q.~Zhao, and S.~Xie, ``Nonnegative matrix and tensor factorizations: An algorithmic perspective,'' \emph{IEEE Signal Process. Mag.}, vol.~31, no.~3, pp. 54--65, 2014.

\bibitem{Gillis2014SPA}
N.~Gillis and S.~A. Vavasis, ``Fast and robust recursive algorithmsfor separable nonnegative matrix factorization,'' \emph{IEEE Trans. Pattern Anal. Mach. Intell.}, vol.~36, no.~4, pp. 698--714, 2014.

\bibitem{fu2015self}
X.~Fu, W.-K. Ma, T.-H. Chan, and J.~M. Bioucas-Dias, ``Self-dictionary sparse regression for hyperspectral unmixing: Greedy pursuit and pure pixel search are related,'' \emph{IEEE J. Sel. Topics Signal Process.}, vol.~9, no.~6, pp. 1128--1141, 2015.

\bibitem{goldstein2014fast}
T.~Goldstein, B.~O'Donoghue, S.~Setzer, and R.~Baraniuk, ``Fast alternating direction optimization methods,'' \emph{SIAM Journal on Imaging Sciences}, vol.~7, no.~3, pp. 1588--1623, 2014.

\bibitem{horn2012matrix}
R.~A. Horn and C.~R. Johnson, \emph{Matrix analysis}.\hskip 1em plus 0.5em minus 0.4em\relax Cambridge university press, 2012.

\bibitem{berman1994nonnegative}
A.~Berman and R.~J. Plemmons, \emph{Nonnegative matrices in the mathematical sciences}.\hskip 1em plus 0.5em minus 0.4em\relax SIAM, 1994.

\bibitem{Wang2012StabilityOM}
Y.-X. Wang and H.~Xu, ``Stability of matrix factorization for collaborative filtering,'' in \emph{Proc. International Conference on Machine Learning (ICML)}.\hskip 1em plus 0.5em minus 0.4em\relax PMLR, 2012.

\bibitem{Wang2004SSIM}
Z.~Wang, A.~Bovik, H.~Sheikh, and E.~Simoncelli, ``Image quality assessment: from error visibility to structural similarity,'' \emph{IEEE Trans. Image Process.}, vol.~13, no.~4, pp. 600--612, 2004.

\bibitem{yun2015ray}
Z.~Yun and M.~F. Iskander, ``Ray tracing for radio propagation modeling: Principles and applications,'' \emph{IEEE Access}, vol.~3, pp. 1089--1100, 2015.

\bibitem{DatasetPaper}
\BIBentryALTinterwordspacing
{\c{C}}.~Yapar, R.~Levie, G.~Kutyniok, and G.~Caire, ``Dataset of pathloss and {ToA} radio maps with localization application,'' \emph{arXiv preprint:2212.11777}, 2022. [Online]. Available: \url{https://arxiv.org/abs/2212.11777}
\BIBentrySTDinterwordspacing

\bibitem{ronneberger2015Unet}
O.~Ronneberger, P.~Fischer, and T.~Brox, ``U-net: Convolutional networks for biomedical image segmentation,'' in \emph{Proc.\ Medical Image Computing and Computer-Assisted Intervention (MICCAI)}, 2015.

\bibitem{compass2008}
T.~King, S.~Kopf, T.~Haenselmann, C.~Lubberger, and W.~Effelsberg, ``{CRAWDAD} dataset mannheim/compass (v. 2008-04-11),'' 2008, [{O}nline]. {A}vailable: \url{https://ieee-dataport.org/open-access/crawdad-mannheimcompass-v-2008-04-11}.

\bibitem{zhou2002covering}
D.-X. Zhou, ``The covering number in learning theory,'' \emph{Journal of Complexity}, vol.~18, no.~3, pp. 739--767, 2002.

\bibitem{shalev2014understanding}
S.~Shalev-Shwartz and S.~Ben-David, \emph{Understanding machine learning: From theory to algorithms}.\hskip 1em plus 0.5em minus 0.4em\relax Cambridge university press, 2014.

\end{thebibliography}

\clearpage
\appendices
\begin{center}
    {\bf Supplementary Material}
\end{center}
\section{Proof of Lemma \ref{lem:boundedset}}
\label{apd:lem_proximal}

As discussed in Section \ref{subsec:recoverability}, the problem to solve becomes \eqref{eqn:ourformulationExplicit} under Assumption \ref{assump_linearDenoiser}, in which the objective function is denoted as $v_{\rm obj}(\{\bm{S}_r,\bm{c}_r\}_{r=1}^R)$.
Denoting the optimal solution to \eqref{eqn:ourformulationExplicit} as $\{\bm{S}_r^\star,\bm{c}_r^\star\}_{r=1}^R$ and the ground truth as $\{\bm{S}_r^\natural,\bm{c}_r^\natural\}_{r=1}^R$, the following inequality holds,
\begin{align}
    &v_{\rm obj}(\{\bm{S}_r^\star,\bm{c}_r^\star\}_{r=1}^R) 
    \leq  v_{\rm obj}(\{\bm{S}_r^\natural,\bm{c}_r^\natural\}_{r=1}^R) \nonumber \\
     = & \frac{\rho}{2}  \sum_{r=1}^R {\bm{s}_r^\natural}^\top\tilde{\bm{Q}}_r (\tilde{\bm{\Lambda}}_r^{-1} - \bm{I} ) \tilde{\bm{Q}}_r^\top {\bm{s}}_r^\natural  + \zeta \sum_{r=1}^{R} {\bm{c}_r^\natural}^\top{\bm{c}}_r^\natural + \left\| \mathbm{O} \oast \mathbm{V}\right\|_{\rm F}^2 \nonumber \\
    \triangleq & v_{\rm obj}^\natural.
    \label{eqn:kappaupper}
\end{align}
On the other hand, $v_{\rm obj}(\{\bm{S}_r,\bm{c}_r\}_{r=1}^R)$ is lower bounded as
\begin{align}
    &v_{\rm obj}(\{\bm{S}_r,\bm{c}_r\}_{r=1}^R)\nonumber \\
    \geq & \bigg\| \mathbm{O} \oast \bigg(\mathbm{Y} -  \sum_{r=1}^R \bm{S}_r^\star \circ {\bm{c}}_r^\star \bigg) \bigg\|_{\rm F}^2  + \frac{\rho}{2} \sum_{r=1}^R {{\bm{s}}_r^\star}^\top\tilde{\bm{Q}}_r (\tilde{\bm{\Lambda}}_r^{-1} - \bm{I} ) \tilde{\bm{Q}}_r^\top {{\bm{s}}_r^\star} \nonumber \\
    \geq & \frac{1}{2} \bigg(\bigg\| \mathbm{O} \oast \Big(\mathbm{Y} -  \sum_{r=1}^R \bm{S}_r^\star \circ {\bm{c}}_r^\star \Big) \bigg\|_{\rm F}  \nonumber \\
    & + \sqrt{\frac{\rho}{2} \sum_{r=1}^R {{\bm{s}}_r^\star}^\top\tilde{\bm{Q}}_r (\tilde{\bm{\Lambda}}_r^{-1} - \bm{I} ) \tilde{\bm{Q}}_r^\top {{\bm{s}}_r^\star}} \: \bigg)^2 \nonumber\\
    \geq & \frac{1}{2}\bigg( \Big\| \mathbm{O} \oast  \sum_{r=1}^R \bm{S}_r^\star \circ {\bm{c}}_r^\star \Big\|_{\rm F} + \sqrt{\frac{\rho}{2} \sum_{r=1}^R {{\bm{s}}_r^\star}^\top\tilde{\bm{Q}}_r (\tilde{\bm{\Lambda}}_r^{-1} - \bm{I} ) \tilde{\bm{Q}}_r^\top {{\bm{s}}_r^\star}} \nonumber \\
    & - \| \mathbm{O} \oast \mathbm{Y}  \|_{\rm F} \bigg)^2.
    \label{eqn:kappalower_s1}
\end{align}
We further define $\bm{s}^\star \in \mathbb{R}^{MNR}$ as the vectorization of $\bm{S}^\star = [{\bm{s}_1^\star},\ldots,{\bm{s}_R^\star}]$, $\bm{Q} \in \mathbb{R}^{MNR\times LR}$ as a block diagonal matrix ${\rm bkdiag}(\tilde{\bm{Q}}_1,\ldots,\tilde{\bm{Q}}_R)$, and $\bm{\Lambda} \in \mathbb{R}^{LR\times LR}$ as a diagonal matrix ${\rm bkdiag} (\tilde{\bm{\Lambda}}_1,\ldots,\tilde{\bm{\Lambda}}_R)$. To simplify notation, we have assumed that ${\bm{W}_r} = \tildebm{Q}_r\tildebm{\Lambda}_r\tildebm{Q}_r$ has rank $L$ for all $r\in[R]$. The proof holds correct even if the ranks are different across different $r$'s. The following inequality can be obtained w.r.t. the $\bm{s}^\star$-related terms,
\begin{align}
    &\bigg(\Big\| \mathbm{O} \oast  \sum_{r=1}^R \bm{S}_r^\star \circ {\bm{c}}_r^\star \Big\|_{\rm F} + \sqrt{\frac{\rho}{2} \sum_{r=1}^R {{\bm{s}}_r^\star}^\top\tilde{\bm{Q}}_r (\tilde{\bm{\Lambda}}_r^{-1} - \bm{I} ) \tilde{\bm{Q}}_r^\top {{\bm{s}}_r^\star}} \bigg)^2\nonumber \\
    &\geq \Big\| \mathbm{O} \oast  \sum_{r=1}^R \bm{S}_r^\star \circ {\bm{c}}_r^\star \Big\|_{\rm F}^2 + \frac{\rho}{2} \sum_{r=1}^R {{\bm{s}}_r^\star}^\top\tilde{\bm{Q}}_r (\tilde{\bm{\Lambda}}_r^{-1} - \bm{I} ) \tilde{\bm{Q}}_r^\top {{\bm{s}}_r^\star} \nonumber \\
    & = {\bm{s}^\star}^\top{\bm \Xi}^\top{\bm \Xi} {\bm{s}^\star} + \frac{\rho}{2} {\bm{s}^{\star}}^\top\bm{Q} (\bm{\Lambda}^{-1} - \bm{I}) \bm{Q}^\top{\bm{s}^{\star}},
    \label{eqn:kappalower_s2}
\end{align}
with $\bm \Xi \in \mathbb{R}^{MNK \times MNR}$ denoting
\begin{align}
    \bm \Xi = \bigg[ \text{diag}(\bm{o})\otimes \bm{c}_1^\star, \text{diag}(\bm{o})\otimes \bm{c}_2^\star, \ldots, \text{diag}(\bm{o})\otimes \bm{c}_R^\star  \bigg],
\end{align}
where $\bm{o}$ is the vectorization of the sampling matrix $\bm{O}$. 

Combining \eqref{eqn:kappaupper}, \eqref{eqn:kappalower_s1} and \eqref{eqn:kappalower_s2}, it can be obtained that
\begin{align}
    &{\bm{s}^\star}^\top( {\bm \Xi}^\top{\bm \Xi} + \frac{\rho}{2} \bm{Q} (\bm{\Lambda}^{-1} - \bm{I}) \bm{Q}^\top) \bm{s}^\star \nonumber \\
    \leq & (\sqrt{2 v_{\rm obj}^\natural} + \|\mathbm{O} \oast \mathbm{Y}  \|_{\rm F})^2.
\end{align}
Considering the constraint $\bm{s}_r \in \mathcal{R}(\tilde{\bm{Q}}_r)$ in \eqref{eqn:ourformulationExplicit} can be reformulated as $\bm{s}_r = \tildebm{Q}_r \bm{t}_r$, and $\bm{Q}^\top\bm{Q} = \bm{I}$ due to $\tildebm{Q}_r^\top\tildebm{Q}_r = \bm{I}$, the above inequality can be rewritten as
\begin{align}
    &{\bm{t}^\star}^\top( \bm{Q}^\top{\bm \Xi}^\top{\bm \Xi}\bm{Q} + \frac{\rho}{2} \bm (\bm{\Lambda}^{-1} - \bm{I}) ) \bm{t}^\star \nonumber \\
    \leq & (\sqrt{2 v_{\rm obj}^\natural} + \|\mathbm{O} \oast \mathbm{Y}  \|_{\rm F})^2,
    \label{eqn:t_bound1}
\end{align}
where $\bm{t}^\star$ denotes $[{\bm{t}_1^\star}^\top,\ldots,{\bm{t}_R^\star}^\top]^\top$. 

Next we will show that $\bm{G} \triangleq  \bm{Q}^\top{\bm \Xi}^\top{\bm \Xi}\bm{Q} + \frac{\rho}{2} \bm (\bm{\Lambda}^{-1} - \bm{I})$ is positive definite (p.d.) under Assumption \ref{assump_linearDenoiser}. Since $\bm G$ is obviously positive semi-definite (p.s.d.), we only need to prove that $\bm{t}^\top\bm{G} \bm{t} \neq 0$ for any non-trivial $\bm t$. We will prove this by contradiction. Let us assume there exists such $\bm t$ so that $\bm{t}^\top\bm{G} \bm t = 0$, then it can be decomposed as
\begin{align}
    \bm{t}^\top\bm G \bm{t} = \sum_{r=1}^R \Big(\bm{t}_r^\top\big(\tilde{\bm{Q}}_r^\top\bm{\Xi}_r^\top\bm{\Xi}_r \tilde{\bm{Q}}_r + \frac{\rho}{2}  (\tilde{\bm{\Lambda}}_r^{-1} - \bm{I}) \big) \bm{t}_r \Big) = 0, \nonumber
\end{align}
which further implies that for any $r\in[R]$,
\begin{subequations}
    \begin{align}
    \bm{t}_r^\top\tilde{\bm{Q}}_r^\top\bm{\Xi}_r^\top\bm{\Xi}_r \tilde{\bm{Q}}_r  \bm{t}_r &= 0, \text{ and } \label{eq:pdProof_1}\\
     \rho \bm{t}_r^\top\bm (\tilde{\bm{\Lambda}}_r^{-1} - \bm{I}) \bm{t}_r &= 0, \label{eq:pdProof_2}
\end{align}    
\end{subequations}
where $\bm{\Xi}_r\in \mathbb{R}^{MNK\times MN}$ is defined as $\text{diag}(\bm o) \otimes \bm{c}_r^\star$---the $r$-th block of $\bm \Xi$.
According to Assumption \ref{assump_linearDenoiser}, the diagonal elements in $\tilde{\bm{\Lambda}}_r$ satisfy $\tilde{\bm{\Lambda}}_r(1,1)\leq 1$ and $\tilde{\bm{\Lambda}}_r(i,i) \in (0,1)$ for $i>1$. If $\tilde{\bm{\Lambda}}_r(1,1)< 1$, then \eqref{eq:pdProof_2} is impossible to hold for any nonzero $\bm t_r$. Therefore, $\bm G$ is p.d. 
Otherwise, if $\tilde{\bm{\Lambda}}_r(1,1) = 1$, then \eqref{eq:pdProof_2} requires $\bm{t}_r(2:L)=0$. This is because $\tildebm{\Lambda}_r^{-1} - \bm I$ is a diagonal matrix, which means that its null space is spanned by $\bm e_1$, where $\bm e_i$ denotes the $i$th unit vector. Taking this result into \eqref{eq:pdProof_1} leads to 
\begin{align}
    \bm{\Xi}_r \tilde{\bm{Q}}_r(:,1)  \bm{t}_r(1)= 0. \label{eq:pdProof_contradiction}
\end{align}
The matrix $\bm{\Xi}_r$‘s column rank is $|\bm \varOmega|$, and its null space is spanned by $\{\bm{e}_i|i\in\OmegaC\}$. To verify this, notice that $\bm \Xi_r= \text{diag}(\bm o) \otimes \bm c_r^\star $, where $\bm o(j)=1$ if $j\in\OmegaVec$, and $\bm o(j)=0$ if $j\in \OmegaC$. Then $\bm \Xi_r \bm e_j$ picks up the $j$th column from $\bm \Xi_r$, which equals $\bm e_j \otimes \bm c_r^\star$ if $j\in \OmegaVec$, and $\bm 0$ if $j\in \OmegaC$. 
For \eqref{eq:pdProof_contradiction} to hold, $\tildebm{Q}_r(:,1)$ should lie in the null space of $\bm{\Xi}_r$. Considering $\OmegaVec$ and $\OmegaC$ are two complementary sets, lying in the null space, i.e., being a linear combination of $\{\bm{e}_i|i\in\OmegaC\}$, requires $\tilde{\bm Q}_r(\OmegaVec,1) = \bm 0$.
However, note that $\tildebm{Q} (:,1)$ is the principal eigenvector of $\bm W_r$. Since $\bm W_r$ is nonnegative and irreducible as stated in Assumption \ref{assump_linearDenoiser}, its principal eigenvector is guaranteed to have all positive elements according to the Perron-Frobenius theorem \cite{horn2012matrix}. This leads to a contradiction since $\tildebm Q_r(\OmegaVec,1) = \bm 0$ cannot hold. Therefore, $\bm G$ is p.d.

Denote the smallest eigenvalue of $\bm G$ as $\lambda_\text{min}$, we can derive from \eqref{eqn:t_bound1} that
\begin{align}
    \|\bm{s}^\star\|_2^2 = \|\bm{t}^\star\|_2^2 \leq \frac{(\sqrt{2v_{\rm obj}^\natural}+\|\mathbm{O}\oast\mathbm{Y}\|_{\rm F})^2}{\lambda_\text{min}}.
    \label{eqn:tball}
\end{align}

Next we will characterize the solution set of $\{\bm{c}_r\}_{r=1}^R$. Considering \eqref{eqn:kappaupper} and that $v_{\rm obj}(\{\bm{S}_r^\star,\bm{c}_r^\star\}_{r=1}^R) \geq \zeta \sum_{r=1}^{R} {{\bm{c}}_r^\star}^\top{{\bm{c}}_r^\star}$, the following can be derived,
\begin{align}
    \| \bm{c}^\star \|_2^2 = \sum_{r=1}^{R} {{\bm{c}}_r^\star}^\top{{\bm{c}}_r^\star} \leq \frac{v_{\rm obj}(\{\bm{S}_r^\star,\bm{c}_r^\star\}_{r=1}^R)}{\zeta} \leq \frac{v_{\rm obj}^\natural}{\zeta},
    \label{eqn:cball}
\end{align}
where $\bm{c}^\star$ denotes the vectorization of $\bm{C}^\star = [\bm{c}_1^\star, \bm{c}_2^\star, \ldots, \bm{c}_R^\star]$. 

\section{Proof of Theorem \ref{thm:recovery}}
\label{apd:theorem_recovery}
In this section, we first present Lemma \ref{lemma3}, which gives the covering number \cite{zhou2002covering} of the optimal solution set to \eqref{eqn:ourformulationExplicit}, an essential component for the proof of Theorem \ref{thm:recovery}. After that, we will provide a detailed proof of Theorem \ref{thm:recovery}.

\subsection{Covering Number of The Solution Set}
Denote the solution set containing $\bm{S}^\star$ and $\bm{C}^\star$ as $\mathbb{S}$ and $\mathbb{K}$, respectively. Based on Lemma \ref{lem:boundedset}, we can now characterize such sets. Recall that we have defined $\beta =  (\sqrt{2v_{\rm obj}^\natural}+\|\mathbm{O}\oast\mathbm{Y}\|_{\rm F})^2/\lambda_\text{min}$ and $\alpha = v_{\rm obj}^\natural / \zeta$. We further define the following:
\begin{align}
    \mathbb{S} &= \{\bm{S} =[\bm{s}_1,\ldots,\bm{s}_R]\in \mathbb{R}^{MN\times R}~| ~\bm{S} \geq \bm 0, \|\bm{S}\|_{\rm F}^2 \leq \beta \}, \nonumber \\
     \mathbb{K} &= \{ \bm{C} = [\bm{c}_1,\ldots,\bm{c}_R] \in \mathbb{R}^{K\times R} ~| ~\bm{C} \geq \bm 0, \| \bm{C} \|_{\rm F}^2 \leq \alpha \}. \nonumber 
\end{align}
In the above, recall that $\|\bm S\|_{\rm F}^2=\sum_{r=1}^R \|\bm S_r\|_{\rm F}^2 = \sum_{r=1}^R \|\bm s_r\|_2^2$ and $\bm s_r={\rm vec}(\bm S_r)$.
In addition, the solution set of the recovered radio map is defined as
\begin{align}
    \mathbb{X}_{\rm sol} = \{ \bm{X} = \bm{C} {\bm{S}}^\top \in \mathbb{R}^{K \times MN} ~| ~\bm{C} \in \mathbb{K}, \bm{S} \in \mathbb{S} \}. \nonumber
\end{align}

\begin{lemma}[Covering Number] \label{lemma3}
The covering number of the $\epsilon_x$-net of
$\mathbb{X}_{\rm sol}$, denoted as $ \text{N}(\mathbb{X}_{\rm sol},\epsilon_x)$, is bounded as
\begin{align}
    \text{N}(\mathbb{X}_{\rm sol},\epsilon_x) \leq \alpha^{KR/2} \beta^{MNR/2}  (\nicefrac{3(\sqrt{\alpha}+\sqrt{\beta})}{\epsilon_x})^{R(K+MN)}.
    \label{eqn:coveringNumber}
\end{align}
\end{lemma}

\begin{IEEEproof}
According to the definition of $\mathbb{S}$, each possible $\bm{s}$ is restricted within a Euclidean ball with radius $\sqrt\beta$, then the covering number of the $\epsilon$-net is bounded by \cite{shalev2014understanding}:
\begin{align}
    \text{N} (\mathbb{S}, \epsilon) \leq (\nicefrac{3\sqrt\beta}{\epsilon})^{MNR}.
\end{align}
Similarly, we have
\begin{align}
    \text{N}(\mathbb{K},\epsilon) \leq (\nicefrac{3\sqrt\alpha}{\epsilon})^{KR}.
\end{align}
Suppose that $\tildebm S$ is from the $\epsilon$-net of $\mathbb{S}$ centered at $\underline{\bm S}$, and $\tildebm C$ is from the $\epsilon$-net of $\mathbb{K}$ centered at $\underline{\bm C}$, then the following holds for any $\tildebm S$ and $\tildebm C$:
\begin{align}
    \| \tildebm{X} - \underline{\bm X} \|_{\rm F} &= \| \tildebm{C} \tildebm{S}^\top- \underline{\bm{C}} \underline{\bm{S}}^\top\|_{\rm F} \nonumber \\
    & = \| \tildebm{C} \tildebm{S}^\top- \tildebm{C} \underline{\bm{S}}^\top+ \tildebm{C} \underline{\bm{S}}^\top- \underline{\bm{C}} \underline{\bm{S}}^\top\|_{\rm F}  \nonumber \\
    & \leq \| \tildebm{C} \tildebm{S}^\top- \tildebm{C} \underline{\bm{S}}^\top\|_{\rm F} + \| \tildebm{C} \underline{\bm{S}}^\top- \underline{\bm{C}} \underline{\bm{S}}^\top\|_{\rm F} \nonumber \\
    & \leq \| \tildebm{C} \|_{\rm F} \|\tildebm{S} - \underline{\bm{S}} \|_{\rm F} + \| \underline{\bm{S}} \|_{\rm F} \|\tildebm{C} - \underline{\bm{C}} \|_{\rm F} \nonumber \\
    & \leq (\sqrt \alpha + \sqrt \beta) \epsilon.
\end{align}
Therefore, $\tildebm{X}=\tildebm C \tildebm S^\top$ is within a $(\sqrt \alpha + \sqrt \beta) \epsilon$-net of $\mathbb{X}$ centered at $\underline{\bm X} =\underline{\bm C} \underline{\bm S}^\top $. By considering all possible combinations of $\epsilon$-nets of $\mathbb{S}$ and $\mathbb{C}$, the covering number of the $(\sqrt\alpha+\sqrt\beta)\epsilon$-net of $\mathbb{X}_{\rm sol}$ can be bounded above by the product of $\text{N} (\mathbb{S}, \epsilon)$ and $\text{N} (\mathbb{K}, \epsilon)$:
\begin{align}
    \text{N}(\mathbb{X}_{\rm sol},(\sqrt\alpha+\sqrt\beta)\epsilon) \leq (\nicefrac{3\sqrt\alpha }{\epsilon})^{KR}(\nicefrac{3\sqrt\beta}{\epsilon})^{MNR},
\end{align}
which equivalently implies \eqref{eqn:coveringNumber} by letting $\epsilon_x = {(\sqrt\alpha+\sqrt\beta)}\epsilon$.
\end{IEEEproof}

\subsection{Proof of The Theorem Statement}
For any optimal solution $\mathbm{X}^\star$ to \eqref{eqn:ourformulationExplicit}, it can be derived that
\begin{align}
    &\frac{1}{\sqrt{MNK}} \left\| {\mathbm{X}^\star} - \mathbm{X}^\natural \right\|_{\rm F} \nonumber \\
    =& \frac{1}{\sqrt{MNK}} \left\| {\mathbm{X}^\star} - \mathbm{Y} + \mathbm{V} \right\|_{\rm F} \nonumber\\
    \leq & \frac{1}{\sqrt{MNK}} \left\| {\mathbm{X}^\star} - \mathbm{Y} \right\|_{\rm F} + \frac{1}{\sqrt{MNK}} \left\| \mathbm{V}\right\|_{\rm F} \nonumber \\
     \leq &\frac{1}{\sqrt{|\bm \varOmega|K}} \left\| \mathbm{O} \oast ({\mathbm{X}^\star} - \mathbm{Y} )\right\|_{\rm F} + \frac{1}{\sqrt{MNK}} \left\| \mathbm{V}\right\|_{\rm F} \nonumber\\
    & + \underbrace{  \Big| \frac{\left\| \mathbm{O} \oast ({\mathbm{X}^\star} - \mathbm{Y} )\right\|_{\rm F}}{\sqrt{|\bm \varOmega|K}}  - \frac{ \left\| {\mathbm{X}^\star} - \mathbm{Y} \right\|_{\rm F}}{\sqrt{MNK}} \Big| }_{\rm{Gap}(\mathbm{X}^{\star},\bm{\varOmega})} \nonumber \\
     \leq& \frac{\sqrt{v_{\rm obj}^\natural}}{\sqrt{|\bm \varOmega|K}} + \frac{\left\| \mathbm{V}\right\|_{\rm F}}{\sqrt{MNK}} + \rm{Gap}(\mathbm{X}^{\star},\bm{\varOmega}) ,\nonumber\\
     \leq & \frac{\sqrt{v_{\rm obj}^\natural}}{\sqrt{|\bm \varOmega|K}} + \frac{\left\| \mathbm{V}\right\|_{\rm F}}{\sqrt{MNK}} + \sup_{\tilde{\mathbm{X}}\in \mathbb{X}_{\rm sol}}\rm{Gap}(\tilde{\mathbm{X}},\bm{\varOmega}) ,
\end{align}
where the second inequality comes from the triangle inequality, and the third inequality is due to the optimality of $\mathbm{X}^\star$; i.e., $\|\mathbm{O}\oast(\mathbm{X}^\star-\mathbm{Y}) \|_{\rm F} \leq v_{\rm obj}(\{\bm{S}_r^\star,\bm{c}_r^\star\}_{r=1}^R) \leq v_{\rm obj}^\natural$.
For conciseness, we will denote $\sup_{\tilde{\mathbm{X}}\in \mathbb{X}_{\rm sol}}\rm{Gap}(\tilde{\mathbm{X}},\bm{\varOmega})$ as $\rm{Gap}^\star(\bm\varOmega)$. In addition, we define the following:
\begin{align}
    {\hattext{Loss}(\mathbm{X})} &= \frac{1}{|\bm\varOmega|K} \sum_{(i,j)\in\bm{\varOmega}} \| \mathbm{Y}(i,j,:) - \mathbm{X}(i,j,:) \|_2^2, \nonumber\\
    {\text{Loss}(\mathbm{X})} &= \frac{1}{MNK} \sum_{i\in[M],j\in[N]} \| \mathbm{Y}(i,j,:) - \mathbm{X}(i,j,:) \|_2^2. \nonumber
\end{align}
Then, $\rm{Gap}(\tilde{\mathbm{X}},\bm{\varOmega})$ can be expressed as $|\sqrt{{\hattext{Loss}(\tilde{\mathbm{X}})}} - \sqrt{{\text{Loss}(\tilde{\mathbm{X}})}}|$.
In addition, the following holds:
\begin{align}
    & \frac{1}{K} \| \mathbm{Y}(i,j,:) - \tilde{\mathbm{X}}(i,j,:) \|_2^2 \nonumber \\
    & \leq \frac{1}{K} (\| \mathbm{Y}(i,j,:)\|_2^2 + \| \tilde{\mathbm{X}}(i,j,:) \|_2^2) \nonumber \\
    & \leq \frac{1}{K} (K \iota^2 + \alpha  \beta  ) \triangleq \xi,
    \label{eqn:errLimit}
\end{align}
where $\iota=\max_{i,j,k} \mathbm{Y}(i,j,k)$. The first inequality is because $\mathbm{Y}$ and $\tilde{\mathbm{X}}$ are both non-negative, the second inequality is because $\tilde{\mathbm{X}}\in \mathbb{X}_{\rm sol}$ 
Taking this upper bound of $ \| \mathbm{Y}(i,j,:) - \tilde{\mathbm{X}}(i,j,:) \|_2^2/K$ into the Serfling's sampling without replacement extension of the Hoeffding's inequality, we have
\begin{align}
    &\text{Pr}( |{\hattext{Loss}(\tilde{\mathbm{X}})} - {\text{Loss}(\tilde{\mathbm{X}})}| \geq \nu)\nonumber\\
    \leq & 2\exp\Big({- \frac{2|\bm{\varOmega}|\nu^2}{(1 - (|\bm\varOmega|-1)/(MN))\xi^2}}\Big).
    \label{eqn:hoeffding_Xstar}
\end{align}
According to the definition of $\rm{Gap}^\star(\bm \varOmega)$, we still need to consider the entire solution set $\mathbb{X}_{\rm sol}$, instead of a specific instance of it as in \eqref{eqn:hoeffding_Xstar}. By denoting the centers of the $\epsilon$-nets of $\mathbb{X}_{\rm sol}$ as $\{\underline{\mathbm{X}}_i\}_{i=1}^{\text{N}(\mathbb{X}_{\rm sol},\epsilon_x)}$, the following can be obtained based on \eqref{eqn:hoeffding_Xstar}
\begin{align}
    &\text{Pr}\Big( \sup_{i\in\text{N}(\mathbb{X}_{\rm sol},\epsilon)}|{\hattext{Loss}(\underline{\mathbm{X}}_i)} - {\text{Loss}(\underline{\mathbm{X}}_i)}| \geq \nu\Big) \nonumber\\ 
    = & \text{Pr}\Big( \bigcup_{i\in\text{N}(\mathbb{X}_{\rm sol},\epsilon)} \Big\{|{\hattext{Loss}(\underline{\mathbm{X}}_i)} - {\text{Loss}(\underline{\mathbm{X}}_i)}| \geq \nu \Big\} \Big)\nonumber\\
    \leq & \sum_{i}^{\text{N}(\mathbb{X}_{\rm sol},\epsilon)} \text{Pr} \Big(  |{\hattext{Loss}(\underline{\mathbm{X}}_i)} - {\text{Loss}(\underline{\mathbm{X}}_i)}| \geq \nu \Big) \nonumber\\
    \leq & 2\text{N}(\mathbb{X}_{\rm sol},\epsilon)\exp\Big({- \frac{2|\bm{\varOmega}|\nu^2}{(1 - (|\bm\varOmega|-1)/(MN))\xi^2}}\Big).
\end{align}
Equivalently, the following inequality holds with probability $1-\delta$,
\begin{align}
    & \sup_{i}|{\hattext{Loss}(\underline{\mathbm{X}}_i)} - {\text{Loss}(\underline{\mathbm{X}}_i)}| \nonumber \\
    \leq & \underbrace{\sqrt{ \Big(\frac{1}{|\bm\varOmega|} - \frac{1}{MN} + \frac{1}{MN|\bm\varOmega|} \Big)  \frac{\xi^2}{2}\log (\frac{2\text{N}(\mathbb{X}_{\rm sol},\epsilon)}{\delta})}}_{\varepsilon(\bm{\varOmega},\delta,\epsilon)} .
\end{align}
For any $\tilde{\mathbm{X}} \in \mathbb{X}_{\rm sol}$, it will be in one of the $\epsilon$-nets with centers $\{\underline{\mathbm{X}}_i\}_{i=1}^{\text{N}(\mathbb{X}_{\rm sol},\epsilon)}$. Denote the center of such $\epsilon$-net as $\underline{\mathbm{X}}_{c}$, then with probability $1-\delta$, the following inequality holds,
\begin{align}
    &\sup_{\tilde{\mathbm{X}}\in \mathbb{X}_{\rm sol}} \left| {\hattext{Loss}(\tilde{\mathbm{X}})} - {\text{Loss}(\tilde{\mathbm{X}})} \right|\nonumber\\
    &=\sup_{\tilde{\mathbm{X}} \in \mathbb{X}_{\rm sol}} \Big| {\hattext{Loss}(\tilde{\mathbm{X}})} - {\hattext{Loss}(\underline{\mathbm{X}}_{c})} + {\hattext{Loss}(\underline{\mathbm{X}}_{c})} \nonumber \\
    &- {\text{Loss}(\tilde{\mathbm{X}})} + {\text{Loss}(\underline{\mathbm{X}}_c)} - {\text{Loss}(\underline{\mathbm{X}}_c)}\Big|\nonumber \\
    & \leq \sup_{\tilde{\mathbm{X}} \in \mathbb{X}_{\rm sol}} \Big\{\Big| {\hattext{Loss}({\mathbm{X}^\star})} - {\hattext{Loss}(\underline{\mathbm{X}}_{c})} \Big| + \Big|{\text{Loss}(\underline{\mathbm{X}}_{c})} \nonumber \\
    &- {\text{Loss}(\tilde{\mathbm{X}})}\Big| +\Big| {\hattext{Loss}(\underline{\mathbm{X}}_c)} - {\text{Loss}(\underline{\mathbm{X}_i)}}\Big| \Big\}\nonumber \\
    & \leq \frac{\epsilon^2}{|\bm{\varOmega}|} + \frac{\epsilon^2}{MN} + \varepsilon(\bm{\varOmega},\delta,\epsilon),
\end{align}
where the last inequality comes from the definition of $\epsilon$-nets. Therefore, with probability $1-\delta$,
\begin{align}
    \text{Gap}^\star(\bm{\varOmega}) &= \text{sup}_{\tilde{\mathbm{X}} \in \mathbb{X}_{\rm sol}}  \left|\sqrt{\hattext{Loss}(\tilde{\mathbm{X}})} - \sqrt{\text{Loss}(\tilde{\mathbm{X}})}\right| \nonumber \\
    &\leq \text{sup}_{\tilde{\mathbm{X}} \in \mathbb{X}_{\rm sol}} \sqrt{\left|  {\hattext{Loss}(\tilde{\mathbm{X}})} - {\text{Loss}(\tilde{\mathbm{X}})}  \right|} \nonumber \\
    & = \sqrt{ \frac{\epsilon^2}{|\bm{\varOmega}|} + \frac{\epsilon^2}{MN} + \varepsilon(\bm{\varOmega},\delta,\epsilon)}.
\end{align}

\section{Proof of Theorem \ref{theorem_convergence}}
\label{Apd:convergence}
We follow the proof in \cite{chan2016plug}, which separates the ADMM updates of Algorithm \ref{alg:admm_pnp} into the following three cases:

\noindent
$\bullet$ \textit{Case 1}: $\triangle_{t+1} \geq \eta\triangle_t$ happens for infinite times, and $\triangle_{t+1} < \eta \triangle_t$ happens for finite times;

\noindent
$\bullet$ \textit{Case 2}: $\triangle_{t+1} \geq \eta\triangle_t$ happens for finite times, and $\triangle_{t+1} < \eta\triangle_t$ happens for infinite times;

\noindent
$\bullet$ \textit{Case 3}: Both $\triangle_{t+1} \geq \eta\triangle_t$ and $\triangle_{t+1} < \eta\triangle_t$ happen for infinite times,

\noindent
in which $\triangle_t$ is the residual defined as in \eqref{eq:triangle}.
In this proof, we will only focus on the convergence property of the non-trivial \textit{Case 1}, as infinite occurrences of $\triangle_{t+1} < \eta\triangle_t$ in \textit{Case 2} already implies the convergence of the variables. In \textit{Case 3}, both conditions $\triangle_{t+1} \geq \eta\triangle_t$ and $\triangle_{t+1} < \eta\triangle_t$ occur for infinite times, making it a combination of \textit{Case 1} and \textit{Case 2}.
Therefore, its convergence is also guaranteed due to the convergence properties of both \textit{Case 1} and \textit{Case 2}; see arguments in \cite{chan2016plug}.

Under \textit{Case 1}, $\rho_{t+1} = \gamma \rho_{t}$ is performed for infinite times. Without loss of generality, we suppose it happens for all iterations. First consider the update of $\bm{Z}_r$ using the denoiser, which corresponds to \eqref{eqn:z_update}. According to Assumption \ref{assump_boundDenoiser}, the following can be derived:
\begin{align}
    &{\| \bm{z}_r^{(t+1)} - (\bm{s}_r^{(t)} + \bm{\psi}_r^{(t)}) \|_2}/{\sqrt{MN}} \nonumber\\
    = & {\| \bm{D}_\sigma (\bm{s}_r^{(t)} + \bm{\psi}_r^{(t)}) - (\bm{s}_r^{(t)} + \bm{\psi}_r^{(t)}) \|_2}/{\sqrt{MN}} \nonumber\\
    \leq & \sigma \sqrt{C} = \sqrt{\frac{\lambda}{\rho_t}} \sqrt{C},
\end{align}
where the last equation comes from the $\bm z_r$-update by setting $\sigma=\sqrt{\lambda/\rho_t}$ in \eqref{eqn:z_update}.

Next, consider the update of $\bm{s}_r$.
In Algorithm \ref{alg:admm_pnp}, the $(t+1)$th update $\{\bm s_r^{(t+1)},\bm c_r^{(t+1)}\}_{r=1}^R$ is obtained by solving \eqref{eqn:admmPrimal} using the HALS method. This method initializes with the $t$th update $\{\bm s_r^{(t)},\bm c_r^{(t)},\bm z_r^{(t+1)},\bm \psi_r^{(t)}\}_{r=1}^R$, and iteratively updates with $J$ steps. Here we only focus on the solution from the $J$th iteration; i.e., $\{\bm s_r^{(t+1,J)},\bm c_r^{(t+1,J)}\}_{r=1}^R$, or $\{ \bm s_r^{(t+1)},\bm c_r^{(t+1)} \}_{r=1}^R$ for brevity.
According to \eqref{eqn:x_update_observed_basic} and \eqref{eqn:x_update_unobserved_basic}, $\bm{s}_r^{(t+1)}$ can also be expressed as $\bm{s}_r^{(t+1)} = \max\{\bm{u}_r^{(t+1)},\bm{0}\}$, {with $\bm{u}_r^{(t+1)}$ denoting the solution to minimizing \eqref{eqn:Lagaragian} for $\bm{s}_r$ without any constraints.} In particular, $\bm{u}_r^{(t+1)}$ is obtained by letting the derivative of \eqref{eqn:Lagaragian} be $\bm 0$, leading to
\begin{align}
    &\frac{1}{ {\rho_t}} \nabla_{\bm{s}_r}  f\bigg( \left\{ \bm s_{r'}^{(t+1,J)},\bm c_{r'}^{(t+1,J)} \right\}_{r'=1}^{r-1},\bm s_r,\bm c_r^{(t+1,J-1)}, \nonumber\\
    & \qquad \qquad \left\{ \bm s_{r'}^{(t+1,J-1)},\bm c_r^{(t+1,J-1)} \right\}_{r'=r+1}^R \bigg) \bigg|_{\bm s_r =  \bm{u}_r^{(t+1)}}  \nonumber \\
    =& - (\bm{u}_r^{(t+1)} - \bm{z}_r^{(t+1)}+ \bm{\psi}_r^{(t)} ).
\end{align} 
For brevity, we will denote the partial gradient w.r.t. $\bm s_r$ as $\nabla_{\bm s_r}f |_{\bm s_r = \bm u_r^{(t+1)}}$.
Adding $\bm s_r^{(t)}$ on both sides and reorganizing the terms, we obtain
\begin{align}
     &\bm{u}_r^{(t+1)} - \bm{s}_r^{(t)}\nonumber \\
     =&- \frac{1}{\rho_t}\nabla_{\bm s_r}f |_{\bm s_r = \bm u_r^{(t+1)}} + \bm{z}_r^{(t+1)} - \bm{s}_r^{(t)} - \bm{\psi}_r^{(t)} \nonumber \\
    =&- \frac{1}{\rho_t}\nabla_{\bm s_r}f |_{\bm s_r = \bm u_r^{(t+1)}}+ \bm{D}_{\sigma_t}( \bm{s}_r^{(t)} + \bm{\psi}_r^{(t)}) - (\bm{s}_r^{(t)} + \bm{\psi}_r^{(t)}).
    \label{eq:u-s}
\end{align}
Then it can be derived that
\begin{align}
    & \| \bm{s}_r^{(t+1)} - \bm{s}_r^{(t)} \|_2 / \sqrt{MN} \nonumber \\
    = & \| \max\{\bm{u}_r^{(t+1)},\bm{0}\} - \bm{s}_r^{(t)} \|_2 / \sqrt{MN} \nonumber \\
    \leq & \| \bm{u}_r^{(t+1)} - \bm{s}_r^{(t)} \|_2 / \sqrt{MN} \nonumber \\
    \leq & {\left\|\nabla_{\bm s_r}f |_{\bm s_r = \bm u_r^{(t+1)}}\right\|_2}/{(\rho_t {\sqrt{MN}})}   \nonumber \\
    & + {\left\| \bm{D}_{\sigma_t}( \bm{s}_r^{(t)} + \bm{\psi}_r^{(t)}) - (\bm{s}_r^{(t)} + \bm{\psi}_r^{(t)}) \right\|_2}/{\sqrt{MN}} \nonumber \\
    \leq & \frac{L}{\rho_t} + \sqrt{\frac{\lambda}{\rho_t}} \sqrt{C},
\end{align}
where the first inequality is due to the nonnegativity of $\bm s_r^{(t)}$,
the second inequality is because of \eqref{eq:u-s} and the triangle inequality, and the last inequality follows from Assumptions \ref{assump_boundDenoiser} and \ref{assump_boundGrad}.

Similarly, {by the update of $\bm \psi_r$ in \eqref{eqn:dualupdate}, we have}
\begin{align}
    & \| \bm{\psi}_r^{(t+1)} \|_2 / \sqrt{MN} \nonumber \\
    = & {\| \bm{\psi}_r^{(t)} + \bm{s}_r^{(t+1)} - \bm{z}_r^{(t+1)}  \|_2}/{\sqrt{MN}} \nonumber\\
    = & \| \bm{\psi}_r^{(t)} + \bm{s}_r^{(t+1)} - (\bm{\psi}_r^{(t)} + \bm{s}_r^{(t)}) \nonumber \\
    & + (\bm{\psi}_r^{(t)} + \bm{s}_r^{(t)}) - \bm{z}_r^{(t+1)}  \|_2/{\sqrt{MN}} \nonumber\\
    \leq & (\|\bm{s}_r^{(t+1)} - \bm{s}_r^{(t)}\|_2 \nonumber \\
    & + \| \bm{\psi}_r^{(t)} + \bm{s}_r^{(t)} - \bm{D}_{\sigma_t}(\bm{\psi}_r^{(t)} + \bm{s}_r^{(t)}) \|_2)/{\sqrt{MN}} \nonumber \\
    \leq & \frac{L}{\rho_t } + 2\sqrt{\frac{\lambda}{\rho_t}} \sqrt{C},
\end{align}
and consequently,
\begin{align}
    & \| \bm{\psi}_r^{(t+1)} - \bm{\psi}_r^{(t)} \|_2 / \sqrt{MN} \nonumber \\
    \leq & (\| \bm{\psi}_r^{(t+1)}\|_2 + \|\bm{\psi}_r^{(t)} \|_2)/{\sqrt{MN}} \nonumber \\
    \leq & {L} (\frac{1}{\rho_t} + \frac{1}{\rho_{t-1}}) + 2\sqrt{{\lambda C}}  (\frac{1}{\sqrt{\rho_t}} + \frac{1}{\sqrt{\rho_{t-1}}}).
\end{align}
Finally, {for $\bm z_r$,} we can derive
\begin{align}
    &{\| \bm{z}_r^{(t+1)} - \bm{z}_r^{(t)} \|_2}/{\sqrt{MN}} \nonumber \\
    =& \|\bm{z}_r^{(t+1)} - (\bm{s}_r^{(t)} + \bm{\psi}_r^{(t)} ) + (\bm{s}_r^{(t)} + \bm{\psi}_r^{(t)} )  \nonumber \\
     & - (\bm{s}_r^{(t-1)} + \bm{\psi}_r^{(t-1)} ) + (\bm{s}_r^{(t-1)} + \bm{\psi}_r^{(t-1)} ) - \bm{z}_r^{(t)}\|_2/{\sqrt{MN}} \nonumber \\
    \leq & \Big(\|\bm{D}_{\sigma_t}(\bm{s}_r^{(t)} + \bm{\psi}_r^{(t)}) - (\bm{s}_r^{(t)} + \bm{\psi}_r^{(t)} )\|_2 \nonumber \\
    & + \|\bm{D}_{\sigma_{t-1}}(\bm{s}_r^{(t-1)} + \bm{\psi}_r^{(t-1)}) - (\bm{s}_r^{(t-1)} + \bm{\psi}_r^{(t-1)} )\|_2 \nonumber \\
    & + \|\bm{s}_r^{(t)} - \bm{s}_r^{(t-1)} \|_2  + \|\bm{\psi}_r^{(t)} - \bm{\psi}_r^{(t-1)} \|_2\Big)/{\sqrt{MN}}\nonumber \\
    \leq & \sqrt{\lambda C}(\frac{1}{\sqrt{\rho_t}} + \frac{4}{\sqrt{\rho_{t-1}}}  + \frac{2}{\sqrt{\rho_{t-2}}}) + {L}(\frac{2}{\rho_{t-1}}+\frac{1}{\rho_{t-2}}).
\end{align}

As $\rho_t$ keeps increasing monotonically, it is not hard to verify that $\{\bm{s}_r^{(t)}\}_{t=1}^\infty$, $\{\bm{z}_r^{(t)}\}_{t=1}^\infty$ and $\{\bm{\psi}_r^{(t)}\}_{t=1}^\infty$ are Cauchy sequences. Therefore, $\{ \bm s_r^{(t)},\bm z_r^{(t)}, \bm \psi_r^{(t)} \}_{r=1}^R$ converges to a fixed point $\{ \bar{\bm{s}}_r,\bar{\bm{z}}_r, \bar{\bm{\psi}}_r \}_{r=1}^R$ with $t\rightarrow \infty$. Notice that the update of $\{\bm c_r\}_{r=1}^R$ corresponds to solving a strictly convex quadratic optimization problem with a unique minimizer, which depends on $\{ {\bm{s}}_r,{\bm{z}}_r, {\bm{\psi}}_r \}_{r=1}^R$. This together with the convergence of $\{ \bm{s}_r,\bm{z}_r, \bm{\psi}_r \}_{r=1}^R$ implies $\{\bm c_r\}_{r=1}^R$ also converges.

\section{Proof of Theorem~\ref{corollary_KKT}}
\label{apd:kkt}
Suppose the linear denoisers satisfy Assumption~\ref{assump_linearDenoiser}, then according to Lemma \ref{lem:proximal}, \eqref{eq:ourformulation} can be expressed explicitly as \eqref{eqn:ourformulationExplicit}. The KKT conditions for \eqref{eqn:ourformulationExplicit} are
\begin{subequations} \label{eq:kkt}
    \begin{align}
        & {\nabla_{\bm s_r}  f(\{\bm s_r,\bm c_r\}_{r=1}^R)}+ \lambda \nabla_{\bm s_r}{ g(\bm s_r)} + \bm \alpha_r + \tildebm Q_r^c \bm \gamma_r = \bm 0, ~ \forall r \in [R],\label{eq:kkt_s_2}\\
        & \nabla_{\bm c_r}{ f(\{\bm s_r,\bm c_r\}_{r=1}^R)} + 2\zeta \bm c_r + \bm \beta_r = \bm 0, ~ \forall r \in [R], \label{eq:kkt_c_2}\\
        & \bm s_r \geq \bm 0, ~ \bm c_r \geq \bm 0, ~ \forall r \in [R], \\
        & \bm \alpha_r \leq \bm 0, ~ \bm \beta_r \leq \bm 0, ~ \forall r \in [R],\\
        & \bm \alpha_r \oast \bm s_r = \bm 0,~\bm \beta_r \oast \bm c_r = \bm 0, ~ \forall r \in [R]\\
        & \bm s_r^T {\tildebm Q_r^c} = \bm 0, ~ \forall r \in [R], \label{eq:kkt_Qrange}
    \end{align}
\end{subequations}
where $g(\bm s_r) = ({\rho}/{2\lambda}) \bm s_r^T\tilde{\bm{Q}}_r (\tilde{\bm{\Lambda}}_r^{-1} - \bm{I} ) \tilde{\bm{Q}}_r^\top  \bm{s}_r$ according to \eqref{eqn:ourformulationExplicit}. Dual variables $\bm \alpha_r$, $\bm \beta_r$ and $\bm \gamma_r$ are associated with the constraints $\bm s_r \geq \bm 0$, $\bm c_r \geq \bm 0$ and $\bm s_r^T \tildebm{Q}^c = \bm 0$, respectively. Note that \eqref{eq:kkt_Qrange} is equivalent to $\bm s_r \in \mathbm{R}(\tildebm{Q}_r)$ in \eqref{eqn:ourformulationExplicit}, since $\tildebm Q_r \perp \tildebm Q_r^c$ according to their definition in Lemma \ref{lem:proximal}. Here, we use \eqref{eq:kkt_Qrange} because its equality format simplifies the expression.

Since Theorem \ref{theorem_convergence} guarantees the convergence of Algorithm \ref{alg:admm_pnp} to the fixed point $\{\barbm s_r,\barbm c_r, \barbm z_r\}_{r=1}^R$, we only need to verify that $\barbm s_r$ and $\barbm c_r$ satisfy the KKT condition in \eqref{eq:kkt} for some $\bm \alpha_r$, $\bm \beta_r$ and $\bm \gamma_r$. First of all, according to Lemma \ref{lem:proximal}, the $\bm z_r$-update  \eqref{eqn:prox_z} is an optimal solution to the following strictly convex problem,
\begin{align}
    &\minimize_{\bm z}~ \lambda g(\bm z_r) + \frac{\rho}{2} \| \barbm s_r -\bm z_r + \barbm \psi_r\|_2^2,  \nonumber \\
    & {\rm subject~to~} \bm z_r^T \tildebm{Q}_r = \bm 0. \label{eq:kkt_zupdate}
\end{align}
Therefore, there exists a dual variable $\bm \kappa_r$ such that the KKT condition for \eqref{eq:kkt_zupdate} is satisfied:
\begin{subequations}
    \begin{align}
    & \lambda \nabla_{ \barbm z_r}{ g(\bm z_r)}\big|_{\bm z_r = \barbm z_r} - \rho \barbm \psi_r + \tildebm Q^c \bm \kappa_r = \bm 0,  \label{eq:kkt_z_update2} \\
    & \barbm z_r^T \tildebm Q_r^c = \bm 0, \label{eq:kkt_z_others}
\end{align}
\end{subequations}
where $\barbm s_r = \barbm z_r$ is used in \eqref{eq:kkt_z_update2} according to Theorem \ref{theorem_convergence}.

Next, consider the subproblem of $\bm s_r$; i.e., solving problem~\eqref{eqn:admmPrimal} w.r.t. $\bm s_r$, recast as follows:
\begin{align}
    & \minimize_{\bm s_r} ~f( \bm s_r,\barbm c_r,\{\barbm s_{r'},\barbm c_{r'}\}_{r'=1,r'\neq r}^R ) + \frac{\rho}{2}\| \bm s_r -\barbm z_r + \barbm \psi_r\|_2, \nonumber \\
    & {\rm subject~to~} \bm s_r \geq \bm 0. \label{eq:kkt_s_subproblem}
\end{align}
Since \eqref{eq:kkt_s_subproblem} is a strictly convex quadratic problem with non-negative constraints, it has a unique optimal solution, given by the updates \eqref{eqn:x_update_observed_basic} and \eqref{eqn:x_update_unobserved_basic}. Moreover, according to the convergence of Algorithm \ref{alg:admm_pnp}, such solution should be the fixed point, i.e., $\barbm s_r$, {once the entire solution sequence has converged}. Therefore, {at the limit of $t=\infty$}, there exists a dual variable $\bm \chi_r$, such that the KKT condition for \eqref{eq:kkt_s_subproblem} is satisfied:
\begin{subequations}\label{eq:kkt_s}
    \begin{align}
        &  \nabla_{\bm s_r}  f\big|_{{\bm s_r} = \barbm s_r} + \rho \barbm \psi_r + \bm \chi_r =\bm 0 , \label{eq:kkt_s_update2} \\ 
        & \barbm s_r \geq \bm 0,~\bm \chi_r \leq \bm 0,~\bm \chi_r \oast \barbm s_r = \bm 0, \label{eq:kkt_s_others}
\end{align}
\end{subequations}
where $\nabla_{\bm s_r} f( \bm s_r,\barbm c_r,\{\barbm s_{r'},\barbm c_{r'}\}_{r'=1,r'\neq r}^R )|_{{\bm s_r} = \barbm s_r}$ is denoted as $\nabla_{\bm s_r}  f\big|_{{\bm s_r} = \barbm s_r}$ for brevity, and \eqref{eq:kkt_s_update2} comes from the stationary condition and $\barbm s_r = \barbm z_r$.
Similarly, for $\{ \barbm c_r\}$, there exists $\bm \omega_r$, such that
\begin{subequations}\label{eq:kkt_c}
    \begin{align}
    &\nabla_{\bm c_r} { f}\big|_{\bm c_r = \barbm c_r} + 2\zeta \barbm c_r + \bm \omega_r = \bm 0, \label{eq:kkt_c_update2} \\ 
     & \barbm c_r \geq \bm 0,~ \bm \omega_r \leq \bm 0,~\bm \omega_r \oast \barbm c_r = \bm 0. \label{eq:kkt_c_others}
\end{align}
\end{subequations}
Thus \eqref{eq:kkt_s_2} can be obtained by adding up \eqref{eq:kkt_s_update2} and \eqref{eq:kkt_z_update2}, and letting $\bm s_r = \barbm s_r$, $\bm \alpha_r = \bm \chi_r$ and $\bm \gamma_r = \bm \kappa_r$; \eqref{eq:kkt_c_2} comes from \eqref{eq:kkt_c_update2} by setting $\bm c_r = \barbm c_r$ and $\bm \beta_r = \bm \omega_r$. The other conditions in \eqref{eq:kkt} follow from \eqref{eq:kkt_z_others}, \eqref{eq:kkt_s_others} and \eqref{eq:kkt_c_others} directly.

\end{document}